\documentclass[acmsmall,screen,nonacm]{acmart}
\acmJournal{TOPS}

\AtBeginDocument{%
  \providecommand\BibTeX{{%
    \normalfont B\kern-0.5em{\scshape i\kern-0.25em b}\kern-0.8em\TeX}}}

\setcopyright{acmcopyright}
\copyrightyear{2023}
\acmYear{2023}
\acmDOI{XXXXXXX.XXXXXXX}

\begin{document}

\title[Maximizing Info Gain in Privacy-Aware AL of Email Anomalies]{Maximizing Information Gain in Privacy-Aware Active Learning of Email Anomalies}

\author{Mu-Huan (Miles) Chung}
\email{mhm.chung@mail.utoronto.ca}
\author{Sharon Li}
\email{sharonsiyuan.li@mail.utoronto.ca}
\author{Jaturong Kongmanee}
\email{jaturong.kongmanee@mail.utoronto.ca}
\author{Lu Wang}
\email{wanglu.wang@mail.utoronto.ca}
\affiliation{%
  \institution{University of Toronto}
  \streetaddress{40 St George St}
  \city{Toronto}
  \state{Ontario}
  \country{Canada}
}

\author{Yuhong Yang}
\email{yuhong.yang@sunlife.com}
\author{Calvin Giang}
\email{calvin.giang@sunlife.com}
\author{Khilan Jerath}
\email{khilan.jerath@sunlife.com}
\author{Abhay Raman}
\email{abhay.raman@sunlife.com}
\affiliation{%
  \institution{Sun Life Financial}
  \streetaddress{1 York St}
  \city{Toronto}
  \country{Canada}
}

\author{David Lie}
\email{david.lie@utoronto.ca}
\author{Mark H. Chignell}
\email{chignell@mie.utoronto.ca}
\affiliation{%
  \institution{University of Toronto}
  \streetaddress{40 St George St}
  \city{Toronto}
  \state{Ontario}
  \country{Canada}
}

\renewcommand{\shortauthors}{M.-H. Chung, et al.}

\begin{abstract}
Redacted emails satisfy most privacy requirements but they make it more difficult to detect anomalous emails that may be indicative of data exfiltration. In this paper we develop an enhanced method of Active Learning using an information gain maximizing heuristic, and we evaluate its effectiveness in a real world setting where only redacted versions of email could be labeled by human analysts due to privacy concerns. In the first case study we examined how Active Learning should be carried out. We found that model performance was best when a single highly skilled (in terms of the labelling task) analyst provided the labels. In the second case study we used confidence ratings to estimate the labeling uncertainty of analysts and then prioritized instances for labeling based on the expected information gain (the difference between model uncertainty and analyst uncertainty) that would be provided by labelling each instance. We found that the information maximization gain heuristic improved model performance over existing sampling methods for Active Learning. Based on the results obtained, we recommend that analysts should be screened, and possibly trained, prior to implementation of Active Learning in cybersecurity applications. We also recommend that the information gain maximizing sample method (based on expert confidence) should be used in early stages of Active Learning, providing that well-calibrated confidence can be obtained. We also note that the expertise of analysts should be assessed prior to Active Learning, as we found that analysts with lower labelling skill had poorly calibrated (over-) confidence in their labels.
\end{abstract}


\received{15 August 2023}

\maketitle

\section{Introduction}
Large organizations are faced with ever increasing cybersecurity threats from a variety of sources. Adversaries, armed with increasingly sophisticated tools are looking for weaknesses that they can exploit. Defending against all possible threats is difficult because of the ingenuity of malicious agents and the many ways in which exploits can be carried out. In the research carried out here, we focus on behavioural analysis as a means of identifying possibly compromised accounts and email anomalies. In this case, hardening the organizational perimeter is insufficient because many threats come either from insiders or from malicious outsiders (or so-called masqueraders) who have obtained valid credentials through various methods such as social engineering. Benign internal users could also leak sensitive information if they are careless, and this can be a major threat, especially when employees are working from home. Thus, organizations have to log and detect any anomalous events so as to respond to (or prevent from) security incidents before further damage is taken. 

Many state-of-the-art in cybersecurity detection approaches have adopted Machine Learning (ML) technologies to support detection tasks. However, while ML algorithms may help improve many detection tasks, they can in turn bring about some new challenges. 

The success of supervised ML technologies depends on high quality training sets. Since there are usually too many logs (logged data) and too few labels in cybersecurity domains, it can be difficult to initiate model training. This situation can get worse due to the fact that training labels in cybersecurity can only be obtained from domain experts (subject matter experts). Experts are expensive, but without expert knowledge, high quality labels cannot be obtained. Furthermore, due to the fast changing characteristics of cybersecurity technologies, possible vulnerabilities, and adversaries’ tactics, previously labeled data and trained models may not be applicable for future detection tasks. Thus, an efficient labeling algorithm that supports continuous model training is needed.

Active Learning (AL) is an interactive labeling method that can provide these required functionalities for cybersecurity domains. The AL training process needs only a small set of labeled data to initiate model training (sometimes cold start without previous labels, begin training with random instances, can also be feasible). The AL model then utilizes a sampling strategy to sort the unlabeled instances, so as to figure out the priority order of instances for which labels are being requested, typically from a human labeler. AL continues this process iteratively, aiming to maximize information gain (model improvement) in each iteration until a stopping criterion (e.g., a sustained level of sufficiently high accuracy) is met. 

Conventional AL models most commonly sort (and query) instances by calculating model uncertainty of each unlabeled instance \cite{holub2008entropy, Joshi2009, Settles2012}. However, under the premise that expertise plays an important role in AL labeling, and that it can affect label reliability significantly, training an AL model that considers only model uncertainty may not be sufficient. Thus, we argue there should be two types of uncertainty that need to be considered under this context: a) model uncertainty and b) human uncertainty. 

\begin{figure}[h!]
    \centering
    \includegraphics[width=0.6\linewidth]{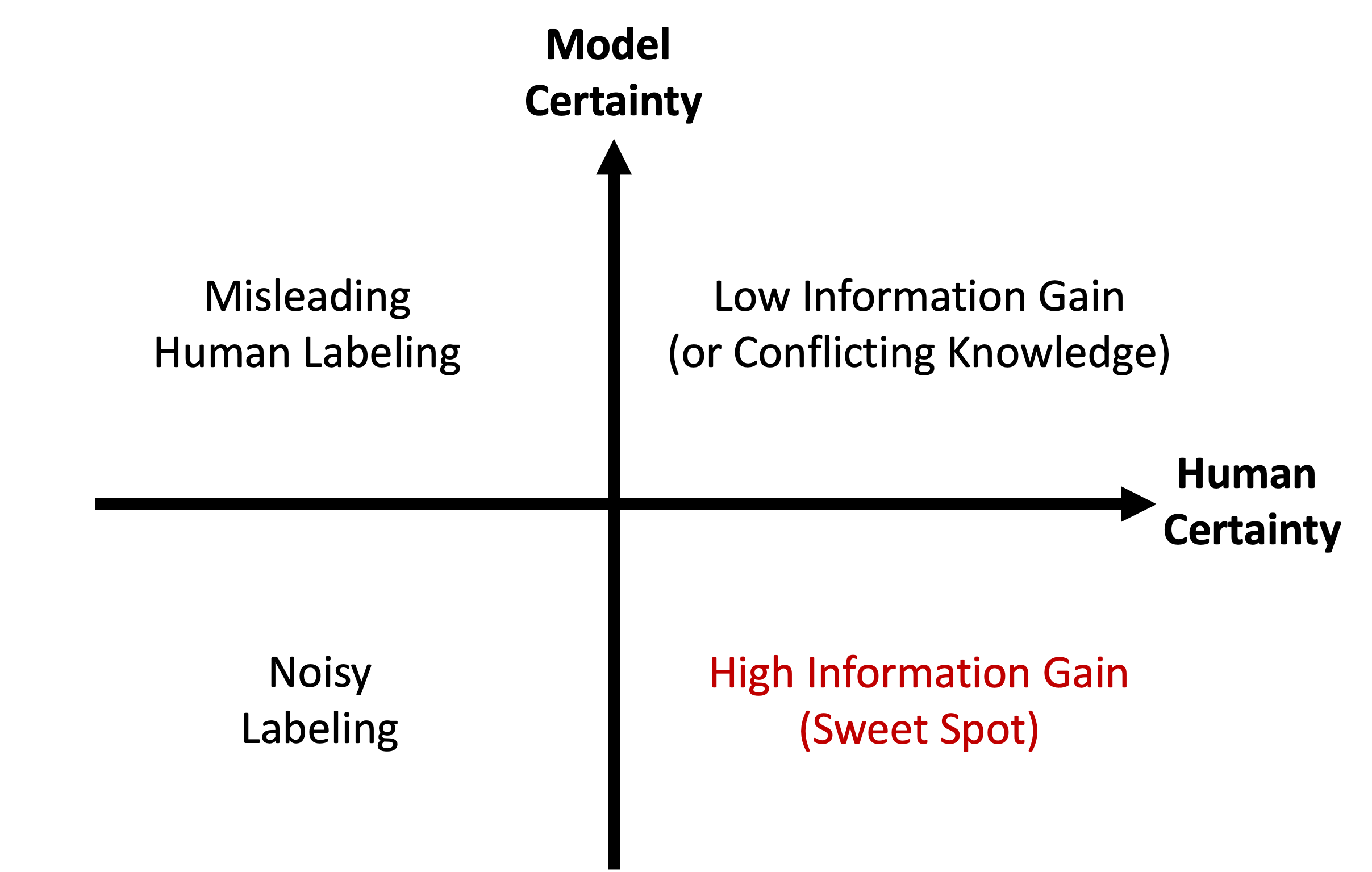}
    \caption{Possible combinations of human and model certainty and uncertainty (highlighting the “sweet” spot where humans are relatively certain and the model is relatively uncertain)}
    \label{fig:1}
\end{figure}

Human uncertainty is the confidence level underlying each expert labeling decision. The relationship between the certainty of information obtained from experts versus model uncertainty is shown schematically in Figure \ref{fig:1}. As shown in Figure \ref{fig:1}, information gain is maximized when the human is certain about the instances being judged, whereas the model has high uncertainty for those instances (lower right quadrant). In such cases, the human can guide the model by providing correct labels, providing high information gain by resolving the model’s uncertainty. In situations where the human has low uncertainty, labeling will generally be noisy, or even misleading for those instances where the model has higher certainty. The high information gain quadrant in the figure is a “sweet spot” where human labeling is useful and where the human’s input can guide the AL process efficiently. 

In contrast, we can regard the top right quadrant in the figure (where the certainty of human and model are both high) as a region where the relatively usefulness of human and machine labeling needs to be assessed based on labeling outcomes. In cases where humans and models have high certainty, and their labels agree, supplementing model predictions with human labels is likely to be unproductive since there is a “low information gain” situation. However, in cases where humans are providing better quality labels (i.e., True model certainty is lower than predicted while the human certainty is relatively accurate), relying on human labels will actually result in useful levels of information gain. 

In the research reported here we focused only on the quadrant in Figure \ref{fig:1} where information gain will be high. The additional quadrant (top right) where human certainty is high will only be useful from an information gain perspective for those cases where the predicted high level of machine certainty is in fact incorrect, as can subsequently be determined by the quality of the labels provided by the model. The question of whether considering the top right quadrant of Figure \ref{fig:1} (in addition to the lower right “information gain” quadrant) can add value to AL over and above using only the information gain quadrant) is a question that we leave to future research. However, we expect that any gains from using the upper right quadrant will be minor relative to the benefits to AL performance from using the “information gain” quadrant where there is likely a discrepancy between human and model uncertainty that is in the human’s favour. 

Based on this information gain approach we amended the AL sampling strategy, and tested the revised AL method using two sets of research questions. Research question set A examined (with case study 1) the situations where experts do not have complete confidence (certainty) in the labels they are assigning, either because their knowledge in the domain is limited, or because the privacy of data does not allow them to have access to all the relevant information (as may often be the case in email anomaly detection). Can AL still work when there is relatively high labeling uncertainty, not just on the ML model side, but also on the side of the human assigning the labels?

\begin{itemize}
    \item \textbf{RQ-1.1}: Do model predictions improve when experts are more confident in their labeling decisions? 
    \item \textbf{RQ-1.2}: Does AL model performance differ when trained by groups versus individuals or dual models (where one person labels for half of the rounds and then the other person labels for the other half of the rounds)?
    \item \textbf{RQ-1.3}: How well do experts agree with each other (how reliable are labels across different analysts)? 
\end{itemize}

Based on the results obtained we suggest the use of an Expert-Derived Information Gain (EDIG) sampling strategy (that focuses on high information gain instances) to improve AL efficiency when training with human experts that may not have complete confidence in the labels they assign. Research question set 2 (case study 2) evaluated the effectiveness of the EDIG strategy in AL, in terms of whether it can help predict and unify expert confidence levels:

\begin{itemize}
    \item \textbf{RQ-2.1}: Does the selection of those query instances (to label), that the human is more likely to be confident about (but the model is uncertain about), lead to faster learning by the model? Alternatively, does consideration of human confidence confer no benefit, and is it sufficient to simply ask the human to label instances that the model is uncertain about?
    \item \textbf{RQ-2.2}: Does the quality of EDIG AL models lead to better a) confidence in labelling; b) label quality (as determined by agreement with ground truth labels)? 
\end{itemize}

In the remainder of this paper we first review the related research in AL. We then summarize the method, process, and results of a case study that examines the application of AL, in a challenging email anomaly detection process, where expert confidence ratings in their assigned labels are also collected. We then describe in detail how EDIG can be implemented in AL, compare EDIG against a previously successful rank-batch mode (RBM) AL strategy, and conduct case study 2 to test whether EDIG can work efficiently with real world data and domain experts. Finally we summarize the results obtained from both quantitative and qualitative (expert self-report) data and answer the research questions that were posed before discussing the implications of the research results for guiding the future use of AL in cybersecurity.

\section{Background}
Labeling cost has always been a major issue for supervised ML model training. In many domains crowdsourcing \cite{schenk2009crowdsourcing, estelles2012towards} helps resolve the issue by recruiting participants to label digits, images, or video captions. Captchas \cite{von2003captcha} area well-known use case, initially implemented as a Turing test, acting as a security check to identify if a user requesting a resource is a human being or a machine. The related ReCaptcha crowdsourcing application was then used to support book digitalization with a 30 million capacity daily \cite{von2008recaptcha}. However, the success in crowdsourcing can hardly be applied horizontally in the domain of cybersecurity. 

Due to the private and sensitive information of cybersecurity data (specifically email anomaly datasets), crowdsourcing of cybersecurity labelling tasks is unacceptable because it requests large scale leaking of confidential data. In addition, the expertise-intensive nature of cybersecurity risk assessment makes it impossible for non-expert participants to provide high quality labels. At the same time, recruiting a large number of cybersecurity analysts is also unreal because of the cost, which will remain high even if external contractors are used and are permitted to have access to sensitive data within an organization. Thus, AL is attractive because it does not need many participants and relies on relatively small numbers of instances to be labelled, and as a result it can be done internally within an organization. 

AL uses a feedback loop to overcome ML training “bottlenecks” \cite{Settles2009}. For many tasks that are naturally easier for human brains, such as object segmentation, video annotation, AL may improve performance by incorporating human knowledge. When labelling involves tasks that people do not perform naturally or intuitively, human input to model training may still be beneficial if the human judges have sufficient experience and expertise to provide useful labels.

\subsection{AL Sampling Strategies}
\begin{figure}[h!]
    \centering
    \includegraphics[width=0.5\linewidth]{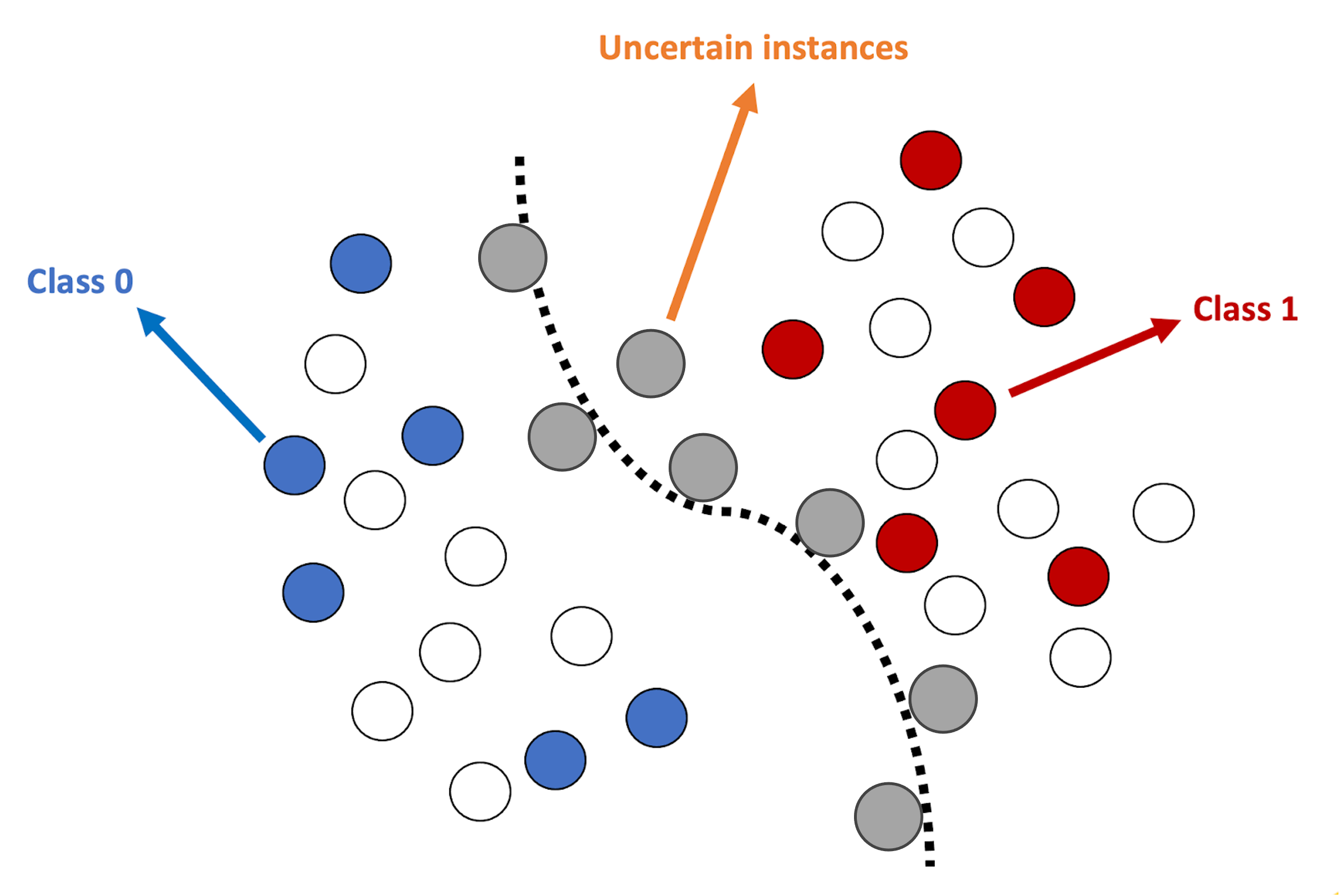}
    \caption{AL implementation in a binary classification task}
    \label{fig:2}
\end{figure}

A sampling strategy is required for an AL model to work. One of the most common strategies is to prioritize instances for labeling based on ML model uncertainty \cite{holub2008entropy, Joshi2009}. Using a type of uncertainty minimization strategy the queue of instances for labels is ordered based on their uncertainty rankings, so that the model can maximize information gain after each labeling judgment. For instance, as shown in Figure \ref{fig:2} (binary classification), the gray bubbles closer to the decision boundary are more likely to be queried in AL because they have higher model uncertainty (due to their proximity to the decision boundary). 

There are also other querying strategies in addition to uncertainty sampling \cite{Settles2009}. In situations such as using probabilistic models, or having multiple models in a pipeline, other methods (e.g., Query-by-Committee, Disagreement Sampling, Expected Model Change, or Estimated Error Reduction, etc.) may better optimize the model with more improvements in each feedback loop. Thus, the decision of the querying strategy is task-oriented and subject to change based on the algorithm underlying the active learner. 

While other querying strategies (e.g., \cite{he2014active}) have been proposed, the uncertainty driven sampling strategy remains a simple and straightforward approach and in the remainder of this paper we explore (within an applied setting) how uncertainty driven sampling for AL can be improved by also considering the uncertainty of the human labeler. 

\subsection{AL Querying Strategies}
AL may operate with different strategies based on the type of data flow and the target task, where there are two major strategies: stream-based and pool-based. The stream-based strategy was first proposed as a sampling strategy to replace random sampling for certain tasks, under the assumption that the unlabeled data acquisition cost is subtle (or free) \cite{atlas1989training, cohn1994improving}. The active learner can decide whether or not to request a label for an upcoming instance or to skip it, based on distributional properties of the data. Since the decision is made for each instance in a stream (or in a sequence/row), the strategy is referred to as stream-based. 

The pool-based strategy focuses on situations where a large set of unlabeled data can be acquired at once \cite{Settles2009}. Pool-based AL typically starts with an initial learner model trained with a small set of labeled data (the initializer). The active learner then query for labels of each instance in a large set of unlabeled data (the pool) \cite{lewis1995sequential}. Once queried an instance is removed from the pool and the model would move on and query the next instance.

The characteristic of requesting one label at a time in both stream-based and pool-based strategies can sometimes be problematic. There are three major issues associated with such querying style \cite{Settles2011,he2014active} which can be summarized as follows: 

\begin{itemize}
    \item \textbf{Redundancy} Selecting one instance at a time, from the most uncertain instance, to the second most uncertain one, and finally the least uncertain instance, (following the order of model uncertainty) can be inefficient. In many cases instances with similar model uncertainty values may be too similar to each other to provide new information.
    \item \textbf{Representativeness} Some instances that have the highest model uncertainty values can be outliers. While they may be uncertain to the model, most likely because they are close to the decision boundary, that part of the decision boundary can be too far from other instance-dense areas, and thus not representative of a sufficient number of cases to provide information that improves overall model performance.
    \item \textbf{Efficiency} Every round of sampling (labelling) is followed by retraining of the model. Retraining the model after each instance is labelled can be very inefficient, especially when a complex, computationally expensive algorithm is used.
\end{itemize}

These issues make stream-based and pool-based strategies incompatible with many modern labeling circumstances. Thus batch mode strategies were proposed that allow the AL model to update only after a batch of multiple instances are collected. The batch mode strategy can reduce the time cost compared with the two previous strategies. The problem is how to find, in each iteration, the best batch of instances to query. 

\subsection{Batch Mode Querying}
Batch mode sampling involves updating the model only after a batch of multiple instances are collected \cite{hoi2006batch, demir2010batch}, so as to reduce time costs relative to other sampling strategies. The concept of batch-mode processing allows the active learner to sample a “batch” of uncertain instances (if using uncertainty sampling) at the same time. This may significantly reduce the time cost of the two traditional scenarios. However, the batch-based scenario sometimes leads to excessive redundancy in a batch, where there may be little to no information gain during an iteration \cite{guo2007discriminative}.

Batch-mode querying strategies often come with some “heuristics”, which are methodologies to select batches that contain instances that are both informative (uncertain to the model, more likely to be representative) and diverse (not redundant).

Many studies have investigated how to optimize batch instance selection. As an example, Brinker used the Support Vector Machine (SVM) algorithm to estimate potential information gain of each unlabeled instance \cite{brinker2003incorporating}. In this way, an AL learner model can choose a diverse set of instances with sufficient information gain. Geometric properties are another good source to help select unlabeled instances into a batch. For example, using cluster centers and geometric distances between instances \cite{xu2003representative, hoi2006batch, hoi2006large, demir2010batch} can help select diverse instances that are still close enough to instance-dense areas. 

Batch mode methods may outperform traditional sampling strategies in terms of efficiency since ML models are only updated after each batch/round and not after each instance has been labeled. They also tend to be more usable since the human judge can label a number of instances at once, rather than having to wait after each instance is labeled for the model to determine what the next query instance in the sampling sequence will be. However, the use of batches means that sampling strategy cannot adapt to information that is learned after each instance is labeled, but only after the batch is completed, potentially compromising model performance and requiring more rounds of AL training. 

\subsection{Rank-Batch Mode Querying}
In order to make batch-based sampling more efficient, Cardoso et al. proposed a rank-batch mode strategy \cite{cardoso2017ranked} focused on improving batch mode sampling strategy based on performance, as well as usability. Cardoso et al. proposed using a scoring system to rank unlabeled instances based on the information already provided by labeled instances. Thus, if the scoring system worked as intended then it should select better (ideally optimal) batches of instances to be labeled. 

Rank-batch mode AL created a new paradigm for applying AL. It considered all three required properties (redundancy, representativeness and efficiency) with a scoring system, so as to select more optimal batches to query the oracle(s). In order to improve usability, the process can be designed so that human labelers can select how large a sample of instances to be labeled should be included in each ranked batch.

\subsection{Over-Sampling Positive Instances }
A key component of AL involves selecting the order of instances to label, using some form of ranking process. Typically, instances ranked higher, and placed earlier, in the priority queue, are closer to the decision boundary, but these will then be instances that are harder to label. Ambiguous labels may not only be difficult to label, but they may also be of less interest to analysts, who will assign higher priority to extreme cases which need to be urgently addressed. Thus there is a tradeoff between labeling ambiguous instances to fine-tune the prediction model, and letting analysts focus on labeling positive instances (e.g., exhibiting data exfiltration or fraud)  that need to be dealt with. 

One of the problems with AL is that it can be frustrating \cite{lee2020empowering} for analysts to have to label cases that the algorithm is uncertain about, since the analyst will likely be uncertain about those cases as well. Thus, we can expect that AL will work best when analysts are able to confidently label the cases that the algorithm finds ambiguous. Presumably in practical situations analyst labeling of ambiguous cases will be limited by how willing analysts are to carry out what is a potentially frustrating task. 

Carcillo et al. proposed querying top instances of predicted positives by the model, combined with a stochastic semi-supervised learning approach to improve AL model performances \cite{carcillo2018streaming}. This approach was found to be able to reduce labeler frustration and can improve AL performance by 5\%, on a highly unbalanced dataset.  

\section{Case Study 1}
In this case study, involving a large financial services company, the effectiveness and efficiency of AL, as well as the human factors of expert-model interactions, were examined. At the request of the company an outbound email anomaly detection task was used that focused on email behaviours that violated company policy (and, in particular, emails that employees sent to their own email address). This type of behavior was judged to be problematic because the company required that employees worked on “locked-down” company computers and employees might circumvent this policy by sending email messages to themselves to covertly transfer data to other (insecure) computers that they used. 

Prior to our research, the detection task was conducted weekly, and manually, by one security analyst, who reviewed emails on a set of dashboards in order to label each email/case as anomalous or not. The dashboards had filtering functionalities, so that the analyst could screen only emails with certain sensitive terms. After filtering, the dashboard set would visualize activities with two scatter plots of all email sizes and counts of all users; two trend plots of a selected user’s past email count and size histories; and a detailed table of the selected user’s emails on that day. The whole process usually took roughly two business days (every week) to complete, which included filtering from roughly 10,000 emails, screening and reviewing over 1000 filtered emails from about 25 to 50 different users who sent numerous and/or sizable emails that contained certain sensitive terms in subject lines or attachment names.

There was also a second stage of investigation conducted by another team that had higher security clearance. The follow-up investigations included reviewing detailed email attributes, email bodies, attachment details, and sometimes talking to the sender’s manager. Unfortunately this process was even less automated and consequently the rate of reviewing each email was much slower.

The whole manual detection process was considered inefficient and prone to misses/false positives. Thus, ML interventions were required so as to improve the detection speed and accuracy. Some previous endeavors using automatic ML (with previous manual detection outcomes as the training/testing datasets) returned only 4\% true anomalies when further investigated in the second stage. Thus, they were deemed ineffective, probably due to the improperly-labeled and imbalanced data issues. As a result, we expected that Active Learning, with its better capability to elicit human knowledge and to guide a ML model, might improve both the manual detection process and ML performance. 

\subsection{Scope and Data}
The dataset used in this study described email instances in terms of  27 columns including the following information:  sender, recipient, subject, attachment, attachment size, sender identifiable data (role, location, hired date, status, etc.), DateTime variables, and whether or not some sensitive terms were mentioned in the subject line or attachment name. We also included two binary variables indicating whether or not an email contained certain sensitive terms (previously used in manual detection tasks) in its subject line (first variable) and attachment names (second variable). We also calculated the similarity of the email address and user name as a proxy for the likelihood that the person was sending an email to one of their other email counts (using Levenshtein distance \cite{levenshtein1966binary}). This similarity was then used as an additional variable.

There were approximately 320,000 rows in the raw, unlabeled dataset, which comprised two weeks of outbound email. This raw dataset was split into eight sets of unlabeled data, with each set being assigned to one of the eight rounds. Data inside successive rounds was ordered by date and time to mimic the ordering of data in a real detection task. A pre-labeled dataset consisting of 200 instances (with ground truth labels supplied by another team who had access to email bodies and attachments) was used to train the initial ML detection model. 

As noted earlier, one factor constraining the labeling task was that the data provided to the analysts did not contain email body and attachment details. The missing (confidential) data was only available to a different team (not studied here) that had higher security clearance and was charged with investigating potential anomalies once they were detected. Thus, in this screening/filtering stage human judges were forced to make labeling decisions based on subject lines, file names, and some user-identifiable information. While these variables should be enough to detect careless exfiltration, they might potentially lead to more anomalies being flagged (“just in case”) in turn leading to a higher rate of false positives). 

\subsection{Methodology}
We based our methodology on that described by Carcillo et al. \cite{carcillo2018streaming} described in section 2.5 but with adjustments to meet the needs of our case study 1. We also used an over-sampling approach where the analysts were shown a high proportion of expected anomalies. However we included a small number of uncertain and random instances in each feedback loop for exploratory purposes. One reason for using the over-sampling approach was that it may be more motivating for human analysts when they are reviewing a significant number of anomalies. 

\subsubsection{Participants}
There were 10 participants in the experiment, representing all the company employees who had the required expertise in the work group that we collaborated with. Participants performed the labeling tasks as part of their work duties. The data resulting from the experiment was collected by a third party within the same company who had no mutual interest nor conflict with the participants. A set of 10 anonymized IDs were generated and assigned to the participants by the third party. Throughout the experiment, the researchers only had access to the IDs and labels, so that they were unable to link model performances with specific participants. In contrast, the third party could link IDs to participants, but could not tell which model was trained by which participant. 

After completing the task the participants were given a choice of whether or not they wished to release their anonymized data for research use. This decision was anonymous and there were no repercussions for people choosing not to complete their labels, or who did not make their data available for research use. This procedure was approved by the University of Toronto ethics review board (Ethics protocol number 42860). 

The labeling task was carried out as part of the work practice of the 10 participants. For each email instance to be labeled, the task was to review the data provided (subject lines, size, file names, etc.), and then determine if the evidence warranted further investigation due to possible data exfiltration activity (inappropriate transfer of confidential data). The participant then provided a label for the instance (anomalous or not anomalous) and moved on to the next email instance in the queue. We used LightGBM \cite{Ke2017} as the underlying active learner model in this case study, for its efficiency and performance in classification tasks. Similar to the strategy used by Carcillo et al. \cite{carcillo2018streaming} we used over-sampling of positive instances (i.e., instances that the ML prediction model was relatively certain were anomalies). 
During each round (day) in the study, each participant labeled 20 instances including 14 instances that were predicted to be anomalies (positive instances) by the ML model. The remaining 6 queries consisted of 3 queries where the model had high uncertainty (UQs), and 3 random queries (RQs).  There were 8 rounds in the study. The amount of data that could be collected was constrained by the amount of time that supervisors at the company thought should be devoted to this work. The raw data was split into 8 unlabeled datasets, so as to mimic a realistic process of daily labeling/detecting tasks. There were no stopping criteria in this case study because we wanted to focus more on the human interactions, rather than looking for a certain level of model performance. 

\subsubsection{Study Design}
Our case study focused on the following constructs/factors:

\begin{itemize}
    \item Individual differences (in terms of experience, expertise, etc.) 
    \item Collaboration effect (when multiple analysts train one model) 
    \item Human uncertainty (i.e., the analyst self-reported confidence level while providing labels, and whether this affects the active learning process) 
    \item Human factors problems noted by analysts during their interaction with the AL system
\end{itemize}

To evaluate possible human factors concerns, as well as answering our proposed research questions, we separated participants into 3 teams (also shown in the following Table \ref{tab:1}, where the different models are labelled with the letters A through F): 

\begin{itemize}
    \item \textbf{Individual}: each participant in the individual team trained (gives feedback to) their own model (B, C, and D) throughout the experiment
    \item \textbf{Swap}: each participant in the swap team trained their own model (E and F) in the first half of the experiment and switched to train a different model (F and E) in the second half of the experiment
    \item \textbf{Group}: participants in the group team trained one model (A) together - in this approach an instance was labeled as 1 (True) if two out of five analysts believed it to be 1 (True)
\end{itemize}

This study design was motivated by the fact that individual differences \cite{gwizdka200712} may impact human-model system performances. For instance, a highly productive computer programmer can do the same amount of work as a larger number of unskilled computer programmers \cite{brooks1995}. The labeling task used here required experience, and expertise, and we wanted to test whether or not differences in skill level between different people would affect the results. We hypothesized that there would be poorer model performance after training when labels were assigned by less skilled analysts (whether working individually or in a group) and we also hypothesized that model performance might worsen when human labelers were swapped between the models halfway through the training process (due to possible effects of inconsistency in the training inputs). One reason for testing the impact of groups is that it has been found to be beneficial In crowdsourcing studies where it may reduce errors \cite{chung2019efficient}, and there has been evidence to suggest that diversity improves decision making in some contexts. Thus on the one hand it could be predicted that labelling by groups might be more beneficial because there are more diverse perspectives within the group. However, on the other hand it could be argued that groups are likely to contain individuals with lower levels of skill which would be likely negatively impact performance relative to a model trained by a single skilled individual. 

\begin{table}[h!]
  \centering
  \caption{Planned group, swap, and individual training teams for AL experiment}
  \label{tab:1}
  \includegraphics[width=0.7\linewidth]{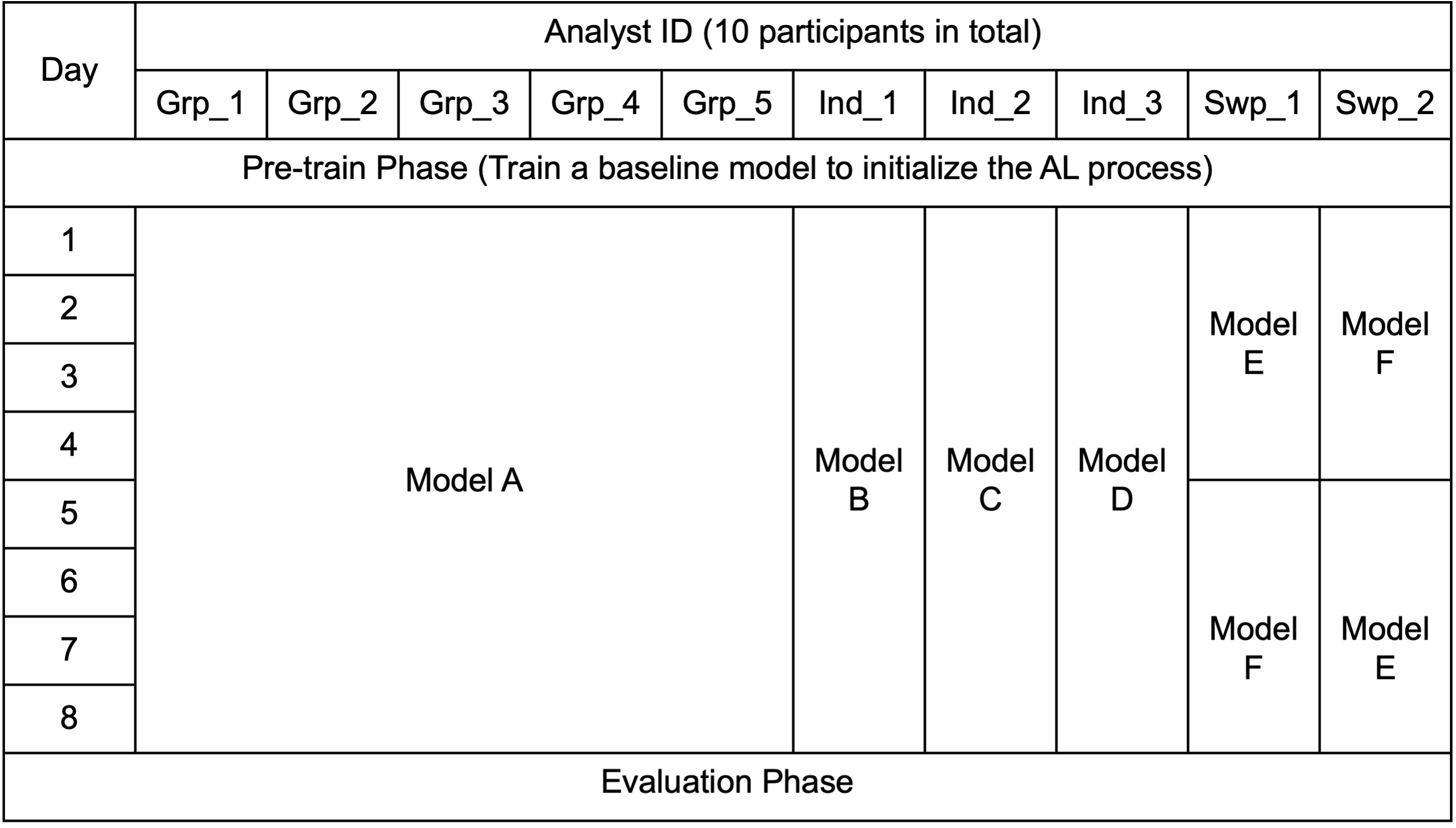}
\end{table}

\begin{figure}[h!]
    \centering
    \includegraphics[width=0.45\linewidth]{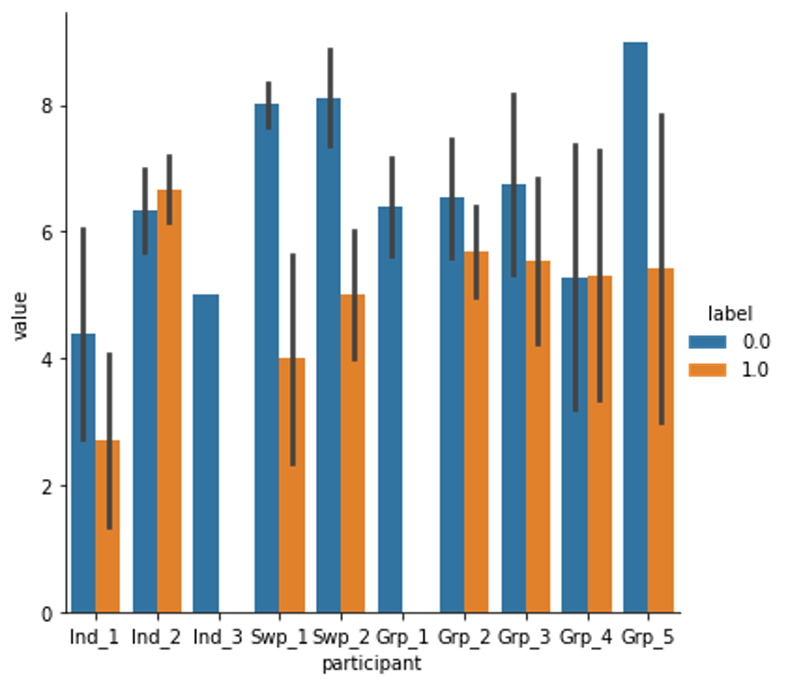}
    \caption{Pre-labeled dataset participant confidence values’ means and error bars for each label class}
    \label{fig:3}
\end{figure}

During the AL process we asked participants to provide their level of confidence with each label that they assigned. This not only allowed us to examine the impact of label uncertainty, but also provided a way to train additional (multiclass) models. The self-reported confidence levels were integers ranging between 0 to 10 (with 0 representing no confidence and 10 representing complete confidence). The average confidence levels and their variations for each class are as shown in Figure \ref{fig:3}. 

As can be seen in Figure \ref{fig:3}, confidence levels were generally higher for 0 (False) instances compared to 1 (True) instances. Thus, a transformation, as shown in the following equation, was applied to convert confidence into a pseudo probability of the instance being 1 (True).

\begin{equation}
  Label_{new} =
    \begin{cases}
      10 - Conf_{i} , \forall Label_{i}=0 \\
      ROUND(\frac{Conf_{i}+10}{2}) , \forall Label_{i}=1
    \end{cases}       
\end{equation}

We based our methodology of data collection and transformation on that described previously by Méndez Méndez et al. They found that collecting participant confidence levels and transforming into probabilities of true positives (in video annotation tasks) was a better way for data annotation, compared with only collecting binary labels \cite{mendez2022eliciting}.

\subsection{Evaluation Schemes and Metrics}
In the last phase of the experiment, we needed a method to evaluate AL results and analyst-model interactions. We used two different ways to evaluate model performance: 

\begin{enumerate}
    \item \textbf{Typical ML model evaluation metrics such as precision and recall values} \\
    Since the dataset was highly imbalanced, we used AUPRC (area under the precision-recall curve), which is preferred over AUROC (area under receiver operating characteristic curve) for its better representation of model performance with imbalanced datasets \cite{saito15}. We also included the F1 score, to balance weightings of recall and precision scores. The test dataset was prepared by another investigation team who had full access to read the email details, including the email body and attachment file content.
    \item \textbf{Amount of expert agreement with the model}\\
    Agreement with the model was determined by assessing the extent to which the over-sampled positive instances were detected as anomalies in each round of the task. We also used Krippendorff’s alpha to evaluate the internal consistency of decisions made within the group \cite{Krippendorff2004, Krippendorff2011}. 
\end{enumerate}

\subsection{RESULTS}
\subsubsection{Model Performance}
The following exposition starts by examining model performance and then addresses the research questions. Since the duration of AL in this study was relatively short due to limited availability of domain analysts, we focus on trends observed in the data and on lessons learned for implementing future AL systems in this type of context. 

Binary model performance (shown in parentheses) was consistently worse than multiclass model performance throughout the study, as shown in Table \ref{tab:2}. Binary labeling (as compared with multiclass labeling) has also performed poorly in other studies (e.g. crowdsourcing studies by \cite{dumitrache2018, mendez2022eliciting}).  It may be unreasonable to expect clearcut binary judgments in ML tasks where experts experience high levels of uncertainty in making the judgment. 

\begin{table}[h!]
  \centering
  \caption{AL performances across different team setups (models) for the multiclass labeling, with equivalent performance for binary labels shown in parentheses}
  \label{tab:2}
  \includegraphics[width=0.35\linewidth]{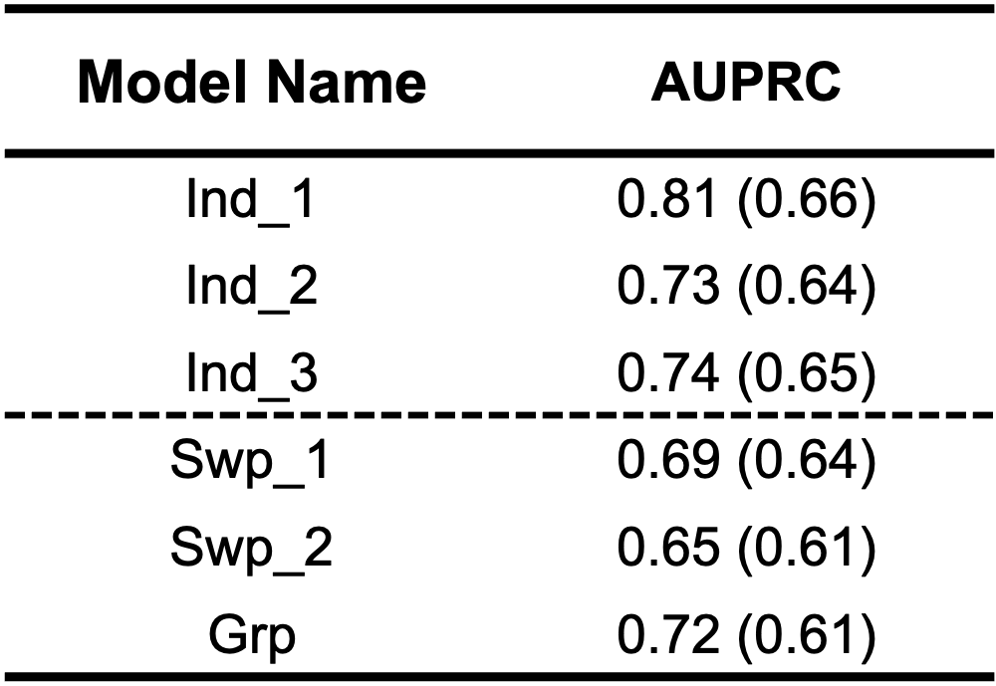}
\end{table}

During training, True anomalies that were labeled by experts as non-anomalies, with confidence ratings of less than 5, were reclassified as true positives, as has been done elsewhere (e.g., \cite{mendez2022eliciting}). Resulting multiclass models performed better than the binary models, likely due to the classes being more balanced during evaluation. When comparing multiclass models, the models trained by the swap and group participants did not outperform models trained by individual participants, with performance being lowest for models trained using a swap condition.

We also implemented evaluation using a second method described in section 3.3 (evaluation method B). In this latter approach we calculated the percentage of the (over-sampled) positive instances being labeled as 1 (True) throughout the case study. This evaluation was possible because 14 out of 20 queried instances of each round were the top anomalies predicted by the AL model. We then assessed whether or not these predictions match analysts decisions/labels in order to assess model performance.

\begin{figure}[h!]
    \centering
    \includegraphics[width=0.45\linewidth]{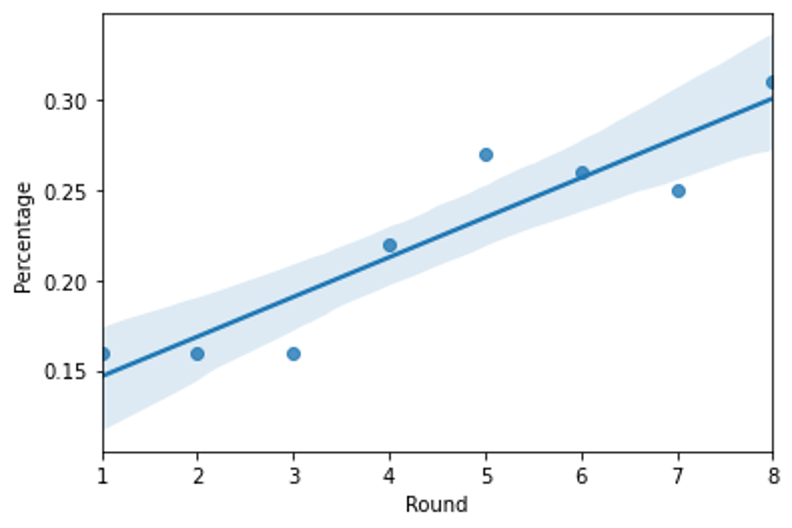}
    \caption{Trend of average 1 (True) label percentages of the first 14 instances in each round}
    \label{fig:4}
\end{figure}

While the average proportion of “True” labels assigned by participants in early rounds was low, an upward trend can be observed (see Figure \ref{fig:4}). The statistical significance of that trend was confirmed using a (non-parametric) Jonckheere Terpstra trend test \cite{jonckheere1954, Terpstra1952}, demonstrating a statistically significant trend (JT=1710.5; p=0.02). This result showed that AL models trained with expert supplied labels were getting more “True” labels over time. As discussed earlier the positive instances included in each round were the model predicted top anomalies. Thus the positive relationship here represented the improvement of model accuracy in terms of matching labeling predictions to expert judgements. 

\subsubsection{Individual Difference and Label Reliability}
\begin{figure}[h!]
    \centering
    \includegraphics[width=0.6\linewidth]{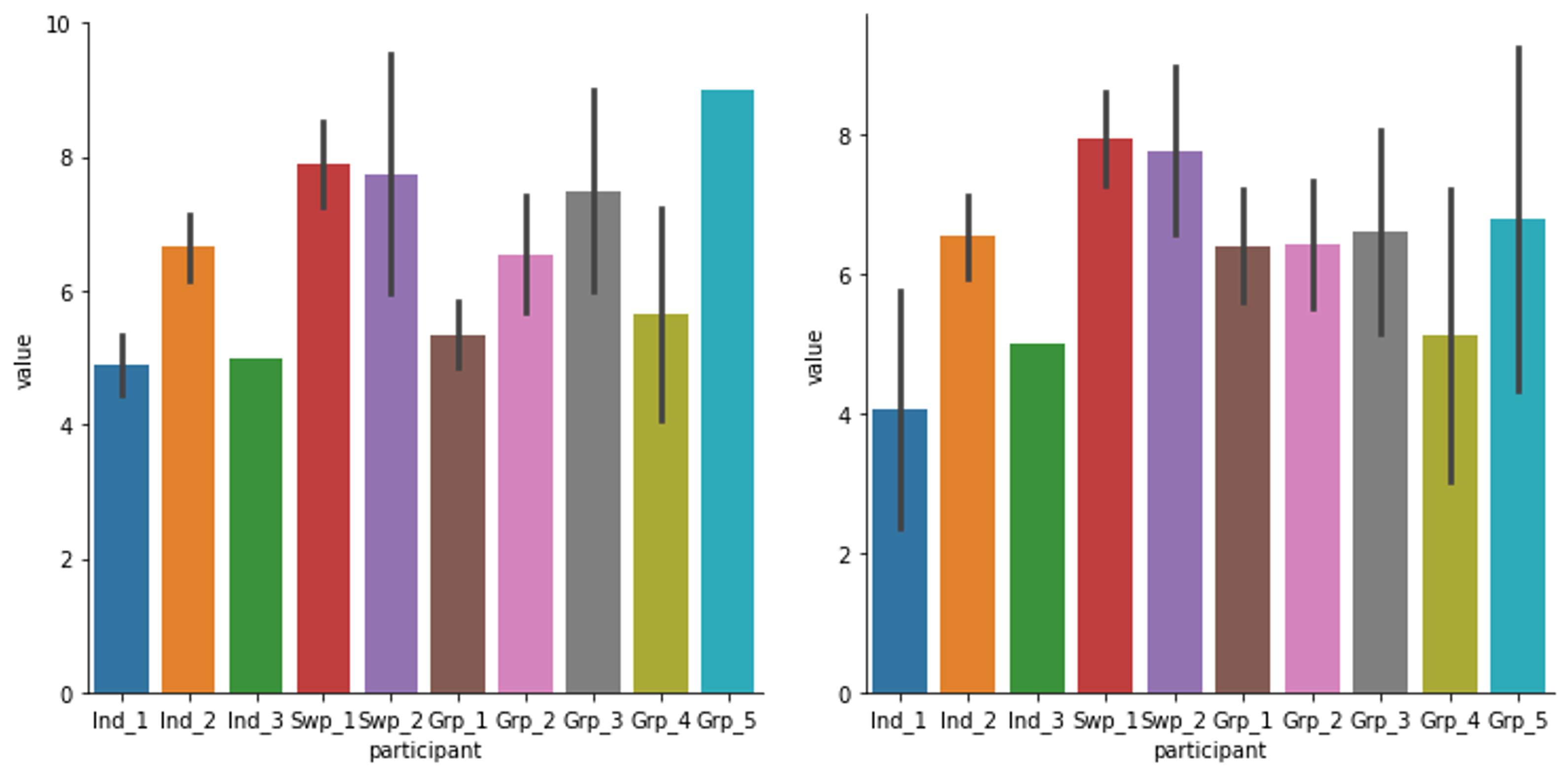}
    \caption{Individual differences in terms of average confidence level values and their variations in round 1 (left panel) and in all rounds (right panel)}
    \label{fig:5}
\end{figure}

As shown in Figure \ref{fig:5} (where the error bars indicate 95\% confidence intervals), participants had differing distributions of confidence ratings with some tending to have more variable ratings, or having higher or lower mean ratings, than others. These differences were observed at the outset (rating differences in the first round of the task are shown in the left panel of Figure \ref{fig:5}), where all participants were queried using the same 20 instances. Strong individual differences were also observed over all the rounds (as shown in the right panel of Figure \ref{fig:5}). 

\begin{figure}[h!]
    \centering
    \includegraphics[width=0.45\linewidth]{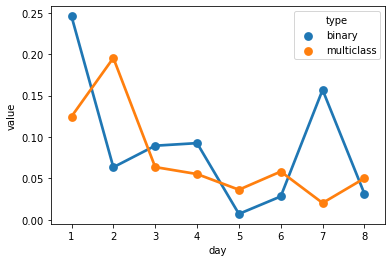}
    \caption{Krippendorff’s alpha values for binary and multiclass model trained with group participants}
    \label{fig:6}
\end{figure}

As a measure of internal consistency within the group, we computed Krippendorff’s alpha values \cite{Krippendorff2004, Krippendorff2011} for the decisions made each day (using the transformed multiclass labels as described earlier) within the group training team (all participants in this team labeled the same 20 instances in every round). Variation in this measure similarity of confidence ratings can be seen in the left panel of Figure \ref{fig:6}, with a generally downward trend in the values being observable. This indicates that analysts in the group training team disagreed with each other more (not less) over time. This may reflect the fact that there was no feedback or consultation between the participants during the study and that motivation may have varied more for some participants than for others, leading to less agreement over time. 

The observed increase in disagreement between experts in the group likely led to reduced label quality, worsening performance of the group training model as compared to the performance of models trained by individual participants. Based on these results we suggest that there should be one or more training sessions with group participants to attain consensus (in practical situations where this technique is to be used). Participants can also be screened based on the results of preliminary labeling of cases, with subsequent AL focusing on high performing individuals, or small subgroups of high performing individuals. This should improve resulting AL model performance relative to current AL practice where groups of judges are used that are heterogeneous with respect to labeling ability.  

\subsubsection{Model and Human Uncertainty}
In terms of label reliability, there exist two types of uncertainties that may affect AL model performances under the set-up of this study: model uncertainty and human uncertainty. Model uncertainty is the measure of how uncertain the model is about some instances, where the most uncertain ones to the model would be queried when using an uncertainty querying strategy; whereas human uncertainty is a latent property that underlies each analyst’s decision, affecting the confidence level as well as the label given.

\begin{figure}[h!]
    \centering
    \includegraphics[width=0.45\linewidth]{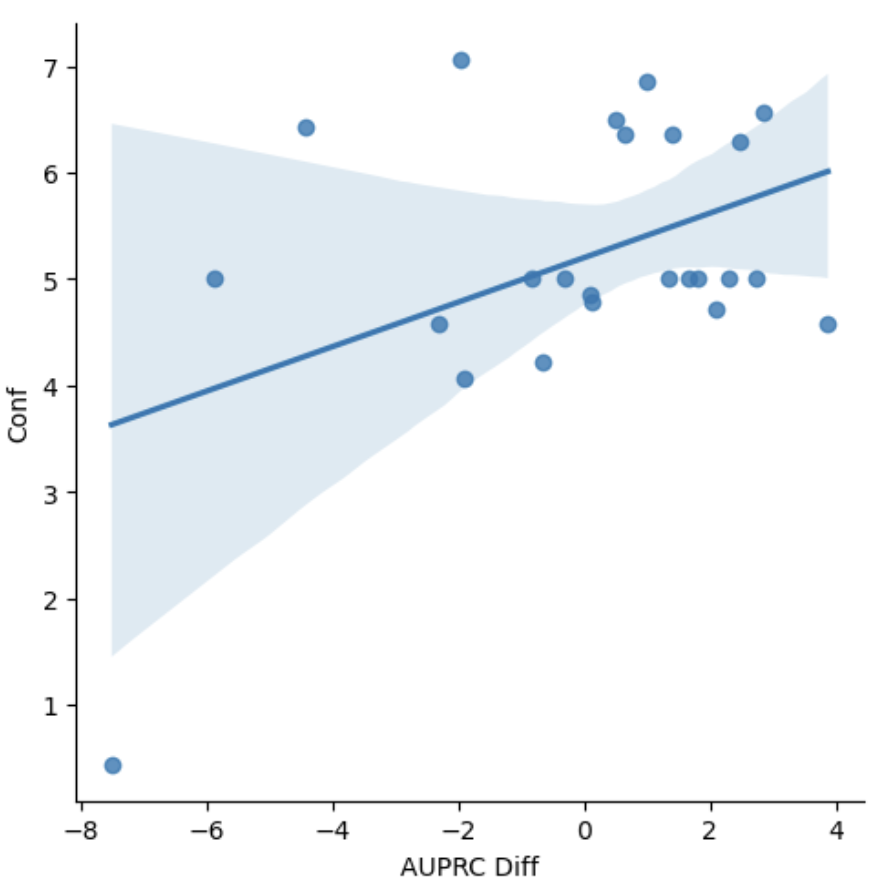}
    \caption{Confidence versus AUPRC changes of models trained by individual analysts}
    \label{fig:7}
\end{figure}

The influence of human uncertainty (from analysts) on model performance can be seen in Figure \ref{tab:2}. Each point in the figures represents the average raw confidence level (Y-axis) and the AUPRC increase/decrease in the resulting model in the following round (X-axis). A statistically significant upward trend ($r=0.44$; $p=0.03$) can be seen with the models trained by the individual participants. Based on this initial result we suggest a hypothesis for future research, that is, there is a positive relationship between analyst confidence and model performance when analysts have a sufficient level of labeling expertise. 

\subsection{Summary of Findings}
In this section we revisit the research questions proposed in section 1, and interpret the effectiveness of AL in a realistic scenario where ground truth labels were not available.  
\newline
\newline
\textbf{RQ-1.1}: Do model predictions improve when experts are more confident in their labeling decisions? 
\newline
\newline
As shown in Figure \ref{fig:7} model performance (for models trained by individual analysts) tended to improve as confidence in labeling increased ($p=0.03$). Thus this research question may be answered in the affirmative. Ideally, if we can trust human confidence judgments with respect to labeling, then confidence judgments provide an upper bound on how much improvement can be expected in active learning, since models will only benefit from training with human-supplied labels if the human labels are in fact reliable. Since it is difficult to establish the reliability of human labels when ground truth labels are not available, confidence might be a useful proxy, in cases where participants have sufficient labeling expertise, for the certainty of human labeling.  
\newline
\newline
\textbf{RQ-1.2}: Does AL model performance differ when trained by groups versus individuals or dual models (where one person labels for half of the rounds and then the other person labels for the other half of the rounds)?  
\newline
\newline
Labeling by a group of people was found to be less effective than labeling based on individuals. This result contradicts typical findings in the crowdsourcing (common AL) literature, where participants possess complete knowledge. In these common AL tasks that do not require expertise, redundancy created by multiple annotators can be helpful as a way to improve label quality and reliability. However, in this case study the participants were only provided with partial information about each email. The different confidence/uncertainty levels between participants in the group team (and also the swap team) may have undermined the overall model improvements. Based on the results obtained, we suggest that in a corporate environment where expertise is required and privacy is a major concern, well-selected and trained individuals can perform better than groups of people with varying amounts of experience/knowledge.
\newline
\newline
\textbf{RQ-1.3}: How well do experts agree with each other (how reliable are labels across different analysts)?
\newline
\newline
Agreement between the participants, as assessed by Krippendorff’s Alpha was relatively low. This doesn’t seem surprising given the difficulty of making labeling decisions in a critical security task, where only partial information is given. Given the strong individual differences that were observed, it seems clear that training sessions are needed to reach a consensus on how to label “correctly”, and to teach participants how to adopt best labeling practices based on the consensus approach identified. Personnel selection may also be needed, so that qualified individuals are recruited for labeling tasks. While the absence of ground truth labels in this case study makes it hard to determine how accurate the participants were in this study, in case study 2 “ground truth” labels from the investigative team were available, and they had access to the complete email information and thus were expected to provide high quality labels.

\section{Case Study 2}
Case study 1 demonstrated that individual labeling was generally better than group labeling in AL in the email anomaly detection application that we considered, and that confidence shows some promise as a proxy for human uncertainty in labeling. In this second case study we trained individual models and we implemented EDIG-based AL using human confidence ratings as proxies for human uncertainty in labeling. In our analysis of the results (reported below) we evaluate how well EDIG based sampling can improve AL efficiency. 

Cardoso et al. \cite{cardoso2017ranked} argued that previous batch mode sampling methods have become obsolete. They proposed a rank-batch mode (RBM) approach \cite{cardoso2017ranked} focused on improving batch mode sampling strategy based on performance, as well as usability.  The RBM approach utilized a scoring system (described in more detail in Section 5.3 below) to rank unlabeled instances according to the information provided by previously labeled instances. This approach is intended not only to help improve the chances of selecting optimal batches, but also to let human labelers select how large a sample of instances to label in each iteration (since the rank of every instance is predetermined by the scoring system). 

\subsection{Expert-Derived Information Gain Strategy}
We based the proposed Expert-Derived Information Gain (EDIG) sampling strategy on the commonly used Rank-Batch Mode (RBM) approach. We modified it so as to find instances for labeling that are in a “sweet-spot” that encompasses query instances that have high uncertainty for the model, but low uncertainty for human judges.   

In order to justify our method for selecting instances within this sweet-spot we first examined, and adapted, a previously proposed scoring function \cite{cardoso2017ranked}. The full form of that proposed RBM scoring system can be seen in equation \ref{eq:5.1}.

\begin{equation}
    \label{eq:5.1}
    Score_u=\alpha*(1-\phi(u,l))+(1-\alpha)*\gamma(u,l),\forall u \in U; l \in L
\end{equation}
\newline 
The scoring equation of the RBM in \ref{eq:5.1} can be written more simply as equation \ref{eq:5.2}:

\begin{equation}
    \label{eq:5.2}
    Score_u=\alpha*diversity+(1-\alpha)*uncertainty
\end{equation}
\newline
The terms labeled as “diversity” and “uncertainty” are further specified by Cardoso et al, and included here for completeness. In this case, $\phi$ denotes the similarity function and $\gamma$ denotes the uncertainty measuring function. Scores are then calculated for each pair of an unlabeled instance and a labeled instance. The $\alpha$ value in equation \ref{eq:5.1} is calculated as $\frac{U}{U+L}$, which is bounded by the sizes of the unlabeled set $U$ and the labeled set $L$. 

This $\alpha$ value is initially set to 1 and gradually decreases to 0 as the AL process progresses (as the pool of unlabeled instances becomes exhausted). Thus the RBM AL model starts with querying diverse instances (where the function is a similarity estimate based on geometric distances) and in later stages looks for instances with higher information gain. In this way the RBM method can dynamically balance instance similarity and uncertainty so as to achieve better AL performance. 

Since the scoring method provided by Cardoso et al. does not take into account human uncertainty, or expert confidence, it will be sub-optimal in cases where human uncertainty varies widely across the space of instances to be labeled. Thus, my proposed EDIG strategy addresses the need for greater efficiency in dynamically controlling the AL selection of batches, by adding an expert confidence term to the scoring function, as shown in equation \ref{eq:5.3}.

\begin{equation}
    \label{eq:5.3}
    Score_u=\alpha*(1-\phi(u,l))+(1-\alpha)*\gamma(u,l)+C(u,l,\beta),\forall u \in U; l \in L
\end{equation}
\newline
The descriptive (simplified) version of equation \ref{eq:5.3} shown below then replaces some terms with labels ("diversity", "uncertainty", and "confidence") that indicate the functional effect of the corresponding terms in equation \ref{eq:5.3}:

\begin{equation}
    \label{eq:5.4}
    Score_u=\alpha*diversity+(1-\alpha)*uncertainty+confidence
\end{equation}
\newline
In this approach, both the diversity and uncertainty properties that past research has found to be crucial in AL literature are retained. We then add the "confidence" term into the equation so as to incorporate the confidence of experts in labeling. Additionally, we added a $\beta$ coefficient to control the relative weighting of confidence. The confidence is calculated in terms of distances from each unlabeled instance to each labeled instance, and then weighted with the confidence input associated with that specific labeled instance.

The first part of equation \ref{eq:5.3} denotes the similarity of each unlabeled instance to each labeled instance. The proximity function can be denoted as equation \ref{eq:5.5}:

\begin{equation}
    \label{eq:5.5}
    \phi(u,l)=\frac{1}{1+\delta(u,l)}
\end{equation}
\newline
where function $\delta$ denotes the distance pair between each unlabeled instance and each labeled instance. In this case We used the Cosine Distance as the proximity function as it has been used frequently by past researchers, and Cardoso et al. found that the cosine distance metric tended to have better performances than the other three distance metrics. 

The second part of equation \ref{eq:5.3} denotes the uncertainty function. Conventionally there are various ways \cite{Settles2012,Danka2018} to calculate uncertainty, such as Least Confident \ref{eq:5.6}, Classification Margin \ref{eq:5.7}, or Classification Entropy \ref{eq:5.8}.

\begin{equation}
    \label{eq:5.6}
    L(x)=1-P(y'|x)
\end{equation}

\begin{equation}
    \label{eq:5.7}
    M(x)=P(y'|x)-P(y"|x)
\end{equation}

\begin{equation}
    \label{eq:5.8}
    H(x)=-\sum_{i}P(x_i|x)logP(x_i|x)
\end{equation}
\newline
In this research we use the Least Confident criterion for uncertainty calculation \ref{eq:5.6} in conformity with recent research practice \cite{tamkin2022active}. The definition/explanation of \ref{eq:5.6} is as follows: in a 3-class classification task, if the classifier predict the probabilities of an unlabeled instance being in either of the 3 class is [0.12, 0.58, 0.30], the uncertainty score of that instance can be calculated as (1-0.58)=0.42. Using the Least Confident function the second part of equation \ref{eq:5.3} can be denoted by the following equation \ref{eq:5.9}:

\begin{equation}
    \label{eq:5.9}
    \alpha(1-\gamma(u,l))=\alpha(1-P(y'|u))
\end{equation}
\newline
where $y'$ represents the highest predicted probability\footnotemark{} within the class of instances that the unlabeled instance $u$ belongs to (where the class membership is predicted by the classifier in use).

\footnotetext{The probabilities used in the calculation of model uncertainty in this chapter are estimated probabilities, as output from a ML classification model. The probability estimations may not represent true probability \cite{guo2017calibration}, as even when accurate overall, probability estimates may not be well-calibrated (e.g., skewed). However, for the purpose of obtaining the “ranking” of samples, in terms of their model uncertainty, the probability estimations can be a good proxy to be used for sorting (e.g., tree-based model output can still be fairly accurate in estimating probabilities \cite{niculescu2005predicting}). As the order of “the most uncertain” samples is an ordinal criterion (e.g., not affected by skew), my scoring function does not require true probability to be valid.}

The last part of equation \ref{eq:5.3} denotes the confidence estimation. This confidence estimation of each unlabeled instance is calculated based on the confidence level provided by the human oracle, weighted by the distance between labeled and unlabeled instance pairs. 

\begin{equation}
    \label{eq:5.10}
    C(u,l,\beta)=\frac{1}{(1+min(\delta(u,l)*\frac{\beta}{\beta+conf_i}))},\forall u \in U; \forall l \in L
\end{equation}
\newline
Equation \ref{eq:5.10} uses weighted confidence with the cosine distance function. The closer an unlabeled instance is to a high confidence instance, the better.

Finally, a recursive hierarchical clustering method (Agglomerative clustering \cite{Mullner2011,Murtagh2014}, was used, prior to scoring, to randomly assign and diversify unlabeled samples (to be selected) into five clusters. For each cluster, only one sample with the highest score would be selected. This approach should help further reduce redundancy (motivated by Figure 3-3 in \cite{Chung2020}, where filtering cases based on the levels of key features prior to training was found to be able to help improve AL performance and efficiency). 

\subsection{EDIG Simulation}
Before applying EDIG to a real world case study, we first tested the method using a simulation approach to see how it compared to the standard AL paradigm. We used four existing datasets from the UCI Machine Learning Data Repository \cite{Frank1998}, and two email anomaly detection datasets collected from the same company as the data analyzed in the preceding chapter, where all of the instances had binary labels and a subset of those had multi-class labels. 

In the simulation we generated the confidence value corresponding to each instance using the following method: 

\begin{itemize}
    \item A separate model (cf. \cite{brinker2003incorporating}) was trained to simulate possible confidence levels by computing the uncertainty of each unlabeled instance based on the labeled instance pool updated by iterations (starting with the second iteration).
    \item The range of confidence is set between [0.3, 0.8] based on the findings in \cite{Chung2020}. Every sample with confidence higher than 0.8 was set to have confidence equals 0.8; whereas if the confidence was lower than 0.3, a sample confidence was set equal to 0.3.
    \item A random float number between 0.2 and -0.2 was added, representing the random error associated with human expert judgements \cite{beckler2018reliability}. 
\end{itemize}

We then randomly split the experiment dataset into a 50\% training set and 50\% testing set with 50 different such splits, so as to better evaluate and compare performance between various supervised learning algorithms \cite{Dietterich1998} in AL that uses both RBM and EDIG. The algorithms used as the classifier (or the uncertainty estimator) in this simulation experiment were as follows: 

\begin{itemize}
    \item k-Nearest Neighbor (kNN) \cite{Cover1967} 
    \item LightGBM (LGBM) \cite{Ke2017}
    \item Support Vector Machine (SVM) \cite{Chang2011}
    \item Naive Bayes (NB) \cite{John2013}
    \item Random Forest (RF) \cite{Breiman2001}
\end{itemize}

We used F1 (equally weighting precision and recall) as the evaluation metric. For each training dataset we used the same AL training criterion. The size of the training set was determined by the smaller of either number of rows of the training set, or 500 instances. In each iteration 5 instances were queried. Thus the maximum number of AL iterations for each dataset was 100.

In this simulation, involving a total of six datasets, we compared how well the new EDIG method compared with the previously used RBM method, in terms of F1-score and how quickly performance improved with respect to the number of iterations.

The results showed performance differences (in F1-scores) across the simulation datasets. We applied the paired Wilcoxon signed rank test \cite{wilcoxon1992individual, rosner2006wilcoxon} (with a 95\% confidence interval to test, with a two-tailed test) to test if performance differences between RBM and EDIG, using the Email\_Binary dataset, were statistically significant. The simulation results suggest that EDIG can yield comparable performance to RBM after 30\% of the total iteration simulated (RF as the underlying algorithm and a $\beta$ value of 0.5). However, this set of results were generated in a simulated AL process, under assumptions that may not necessarily hold true in real world situations. The assumptions can be summarized as follows:

\begin{itemize}
    \item \textbf{Assumption 1}: Model uncertainty is correlated with human uncertainty.
    \item \textbf{Assumption 2}: Human uncertainty will show a higher correlation with label quality (or correctness) than with model uncertainty (although the correlation with model uncertainty will also exist, as noted in assumption 1).
    \item \textbf{Assumption 3}: Shorter geometric distance between two instances (in a high dimensional space) implies that the amount of confidence associated with labeling on the points should be simply to the corresponding level of confidence for the other point.
    \item \textbf{Assumption 4}: Human labelers should be reasonably well calibrated in terms of how well their ratings of confidence in their labels are related to measurable uncertainty in their labels.
\end{itemize}

Since these assumptions needed to be tested with real conditions and with actual human labelers, so as to confirm the usability and feasibility of EDIG in expertise-intense domains, we report below on a study that was carried out to address this need.

\subsection{Scope and Data}
We conducted a case study using a real word dataset (Email\_Binary dataset) to examine the EDIG performances as well as its impact to human labelers, using RF as the underlying ML algorithm with a $\beta$ value of 0.5.\

The dataset used in this study was comprised of 43 features (columns) that represented each email, including sender, recipient, subject, attachment, attachment size, sender identifiable data (role, location, hired date, status, etc.), DateTime variables, and whether or not some sensitive terms were mentioned in the subject line or attachment name (as used in the previous case study 1). In addition, the continuous variables (such as count of sensitive terms in subject, size of attachments in total, number of attachments, etc.) of each email were compared with the sender’s history (within the past 6 months before the email under consideration was sent). 

The dataset was a filtered subset from a streaming email log (collecting around 300 thousand emails every month). The filtered data then was sent to an investigation team at the company where our research was conducted and labeled with detailed assessment. 60\% of the resulting, fully labeled cases in the dataset were anomalies, and more than 80\% of those anomalies were considered minor violations of company policy. Each of the emails, including the attached files, were viewed and investigated thoroughly, providing us with high quality ground truth labels. The data provided to the analysts did not contain email body and attachment details (as in case study 1), because that extra data was only released to the investigation team, who had higher security clearance. 

\subsection{Methodology}
This study implemented an AL process, comparing RBM (the baseline) and EDIG (our proposed approach), in terms of their performance and the confidence that human analysts had in the labels that they provided. The models were updated based on the labels provided by the analysts. 

\subsubsection{Participants}
There were 10 participants in the experiment, representing all the company employees who had the required expertise in the work group that we collaborated with. Participants performed the labeling tasks as part of their work duties. The experimental data were collected using an email labeling dashboard (details about this dashboard to be provided in the following subsection). Participants created their own login account, and communications with the back-end were encrypted. Data were anonymized and even the researchers did not have information about which data belonged to which person. 

After completing the task the participants were given a choice of whether or not they wished to release their anonymized data for research use. This decision was anonymous and there were no repercussions for people choosing not to complete their labels, or who did not make their data available for research use. However, all eligible participants agreed to make their data available for research use. The experimental procedure (protocol) was approved by the University of Toronto ethics review board (Ethics protocol number 43665). 

\subsection{Results}
\subsubsection{Model Performance Comparison (EDIG vs RBM)}
\begin{figure}[h!]
    \centering
    \includegraphics[width=0.45\linewidth]{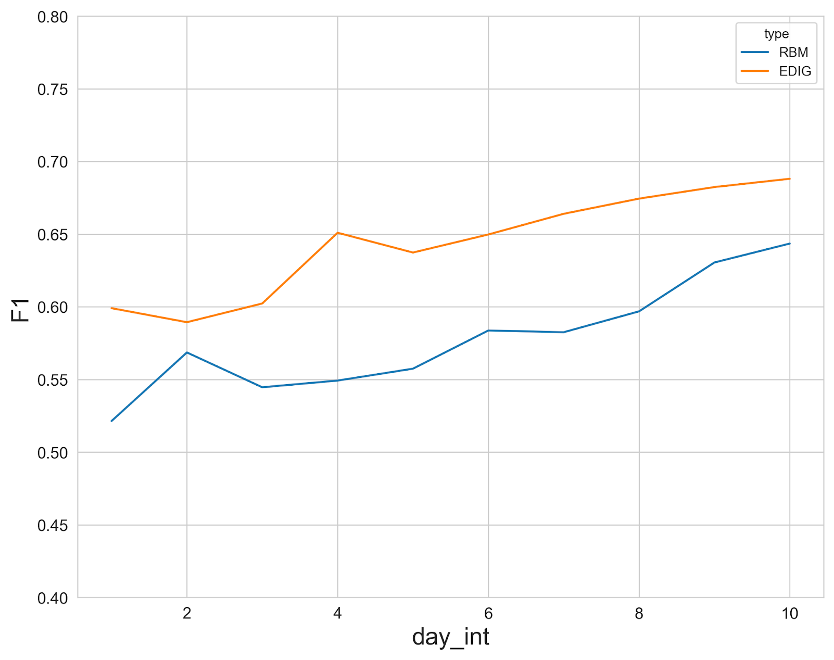}
    \caption{EDIG and RBM performance comparison in F1 score}
    \label{fig:8}
\end{figure}

The first set of results generated for this case study regards model performance. As can be seen in Figure \ref{fig:8}, EDIG AL showed steady improvement, and better F1 performance, compared with RBM AL. The EDIG version of AL was significantly better in terms of F1 than the RBM version as indicated by a Mann-Whitney U test ($U=8.0$; $n1=n2=10$; $p=0.002$). In general, the F1 value for EDIG was around 10\% higher than the corresponding F1 value for RBM, showing that there was an overall improvement in this balanced measure of recall and precision.

\begin{figure}[h!]
    \centering
    \includegraphics[width=0.65\linewidth]{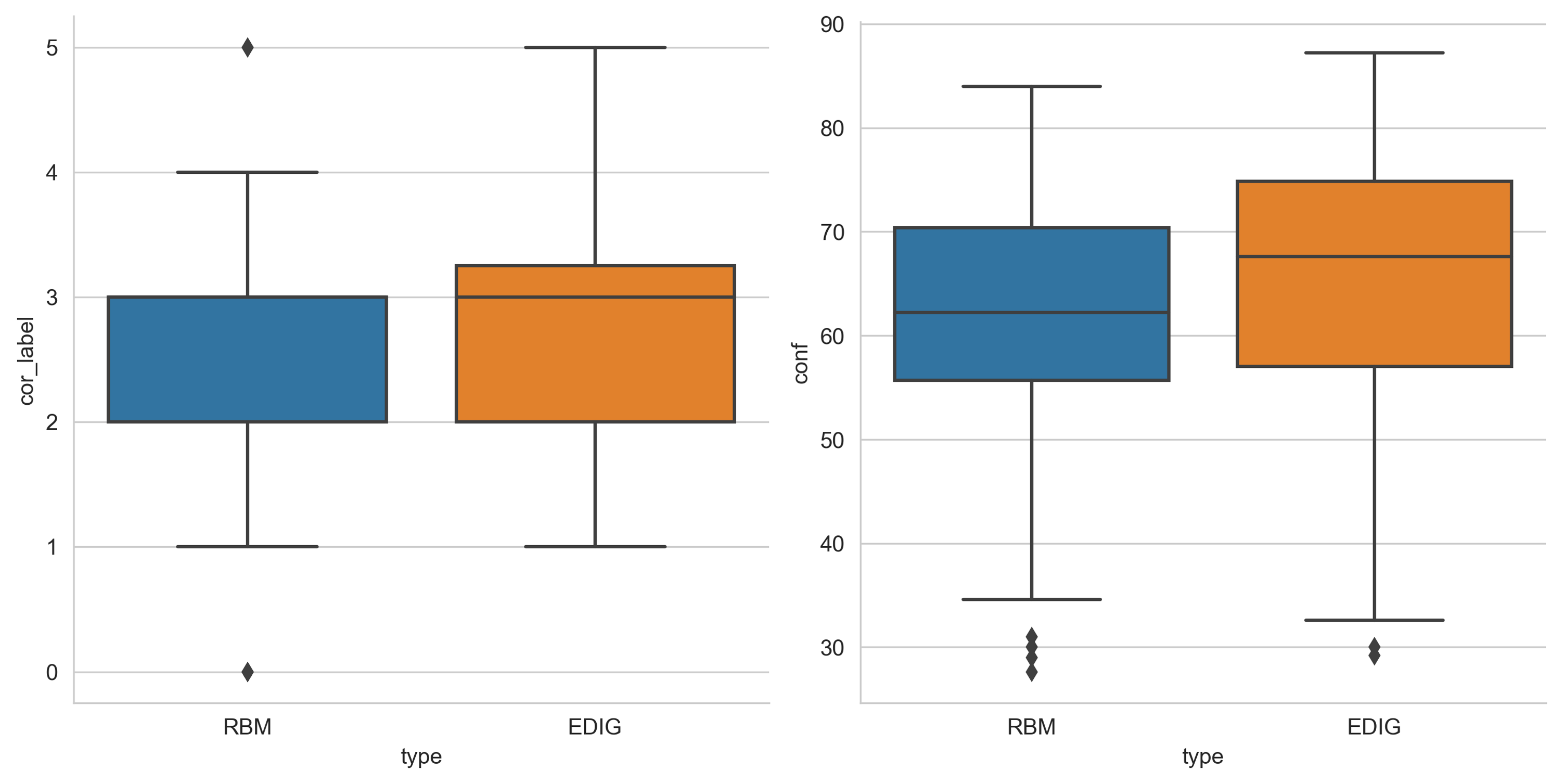}
    \caption{EDIG and RBM comparison in correctly provided label count (left panel); analyst self-reported confidence (right panel)}
    \label{fig:9}
\end{figure}

Figure \ref{fig:8} shows how well the machine performed after different amounts of training (in terms of F1). In contrast, Figure \ref{fig:9} (left panel) shows how well the human is performing in terms of the correctness of the labels that were assigned. Human performance, as assessed using counts of correctly provided labels, was also higher with EDIG selection of instances to be labeled than RBM selection but the relationship was non-significant ($U=2735.5$; $n1=n2=100$; $p=0.22$), possibly because of the constrained sample size (number of available analysts).

As can be seen in Figure \ref{fig:9} (right panel), average expert self-reported confidence for the EDIG models tended to be higher than RBM models (with borderline statistical significance as tested using the Mann-Whitney U test ($U=4293.5$; $n1=n1=100$; $p=0.084$). Given that there were only 10 participants and 10 iterations in this study, it is likely that the advantage of EDIG over RBM would have been statistically significant ($p<.05$) with a larger sample.

\begin{figure}[h!]
    \centering
    \includegraphics[width=0.5\linewidth]{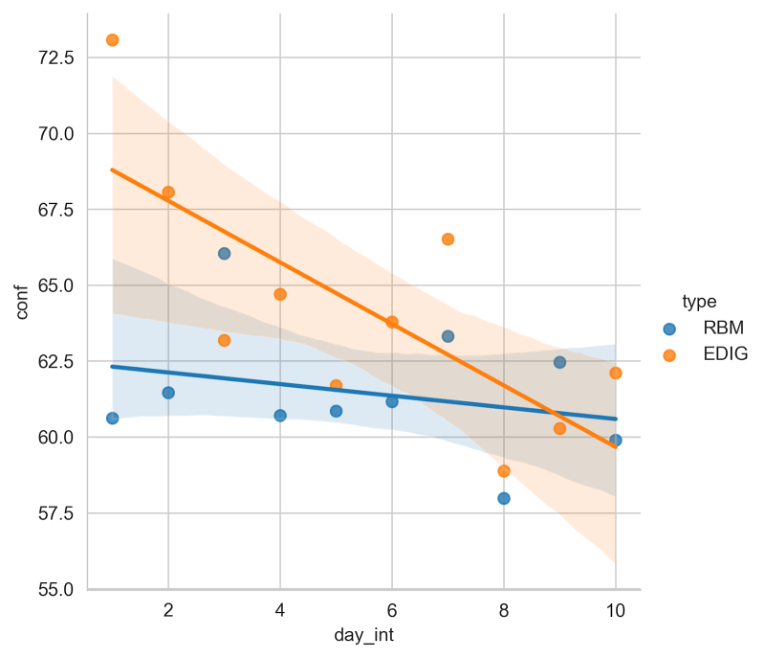}
    \caption{Average confidence levels of RBM and EDIG throughout the case study}
    \label{fig:10}
\end{figure}

The reason why the label accuracy and confidence of EDIG were not significantly greater than the corresponding measures for RBM might be due to increasing difficulty in labeling instances over time as the AL process looks for those cases that the model remains uncertain about (likely close to the decision boundary). In this state of higher uncertainty EDIG guided selection of instances to be labeled no longer conferred an advantage. However, EDIG use was advantageous in early rounds of labeling, where the labeling task was not as difficult. In the first five rounds of this experiment (as shwon in Figure \ref{fig:10}), the average confidence for the EDIG models was (statistically) significantly higher than the corresponding confidence for the RBM models ($U=945.0$; $n1=n2=50$; $p=0.036$). 

There was a significant negative correlation between round (day\_int) and level of confidence ($r = -0.74$, $p = 0.014$) in the EDIG condition, indicating a steadily decreasing confidence when queried by the EDIG models over time. In contrast, the average confidence levels in RBM had a slight downward trend but that trend was not statistically significant. The levels of confidence in both models converged at around 60\% by the end of the case study (round 10). Thus it seems that the EDIG heuristic may work best in the early rounds of AL. 

In both RBM and EDIG scoring functions, the weight of the “model uncertainty” part in equations \ref{eq:5.1} and \ref{eq:5.3} continued to increase due to the effect of the value, potentially making queries more difficult over time for the analysts. While the proposed EDIG method intentionally made early rounds less difficult for analysts, using the “confidence” part in equation \ref{eq:5.3}, in later rounds it appears to have been more difficult to identify instances for labeling that had both high model uncertainty and relative high confidence, as assessed by the human analyst.

This result suggests that there should be a stopping criterion for EDIG, because as the model learns, and cases to be labeled become more uncertain, the EDIG criterion will no longer be helpful, because the experts will then have reduced confidence in their ability to label what may be highly uncertain cases. One idea in this regard is to use a function, where there is a threshold in expert-self reported confidence, below which the AL process is stopped. Alternatively, the labeling interface may provide  an option for the expert to terminate the AL when the labeling task is perceived as being too difficult. However this latter strategy may not work if the expert is “skilled and unaware”, i.e., is providing good labels even though she is not confident in the labels being applied.

\subsubsection{Human Performance Comparison}
\begin{figure}[h!]
    \centering
    \includegraphics[width=0.5\linewidth]{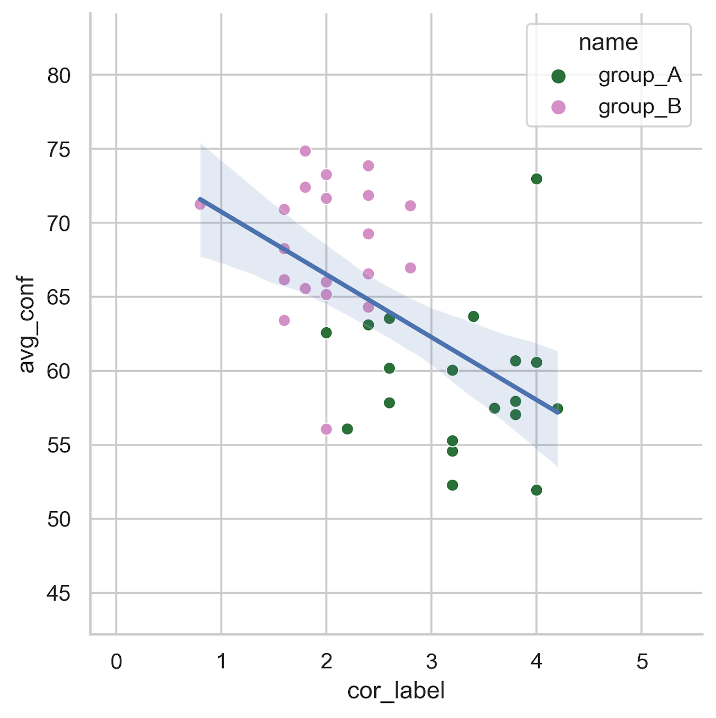}
    \caption{Confidence and label accuracy between group A and group B}
    \label{fig:11}
\end{figure}

As discussed in the previous case study 1, there appeared to be differences in skill level between the labelers. A similar phenomenon was observed when we plotted the final round EDIG performances (in F1 scores) of all participants and applied a selection criterion (Figure \ref{fig:11}). Since the final averaged EDIG performance had an F1 around 0.70 (Figure \ref{fig:8}), we divided the participants into two groups: group A ($>=0.70$ F1) and group B ($<0.70$ F1). After dividing the participants into more and less skilled groups we found a possible Dunning-Kruger effect. The better performing group (group A) provided more correct labels ($U=373.0$; $n1=n2=20$; $p<0.001$), but with lower average self-reported confidence ($U=34.0$; $n1=n2=20$; $p<0.001$).

A Pearson's correlation coefficient was computed to assess the relationship between correct labels provided and average confidence (with the data pooled across both groups). There was a significant negative correlation ($r=-0.55$, $p<0.001$), showing that people who reported greater confidence also tended to provide less accurate labels, consistent with the expectation of a Dunning-Kruger effect. 

\begin{figure}[h!]
    \centering
    \includegraphics[width=0.5\linewidth]{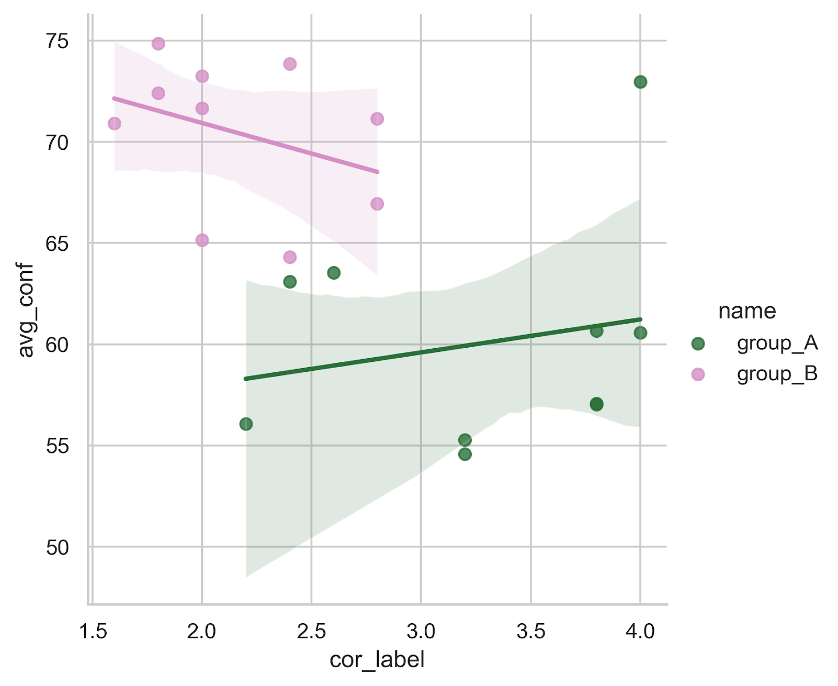}
    \caption{Trendlines of EDIG confidence against correct label counts between group A and group B participants}
    \label{fig:12}
\end{figure}

Figure \ref{fig:12} shows the relationship between confidence and accuracy considered separately within the two groups. Group A participants tended to provide more accurate labels when they were confident ($slope=0.33$; $N=10$), while greater confidence in group B participants tended to lead to less accurate labels, with the correlation between confidence and accuracy being negative ($slope=-0.60$; $N=10$). The difference in the correlations across the two groups was borderline (statistically) significant, as tested with Fisher's z test ($z=1.94$; $p=0.052$). Group A had better calibration of confidence in labeling as reflected in their position correlation between confidence ratings and label accuracy (vs. the negative correlation found for group B).

\subsubsection{Label Quality/Reliability}
As can be seen in Figure \ref{fig:10} there appeared to be a downward trend in self-reported confidence values over time. This may have resulted from a secondary relationship created by the EDIG scoring function, since the weight of the “model uncertainty” part would increase with time. 

In addition to confidence ratings, as can be seen in Figure 5-13, the total number of correct labels of each group tended to decrease over time for group A ($r=-0.19$; $p=0.137$), but tended to increase over time for Group B ($r=0.16$; $p=0.325$), although neither of these trends were statistically significant. The Overall label accuracy of group A members was still higher than group B, and the number before round 10 still tended to favor group A, even though the number of correct labels supplied started dropping after that point for Group A (likely due to the labeling task becoming more difficult, with an associated drop in confidence as shown in Figures \ref{fig:10} and \ref{fig:12}). 

\begin{figure}[h!]
    \centering
    \includegraphics[width=0.5\linewidth]{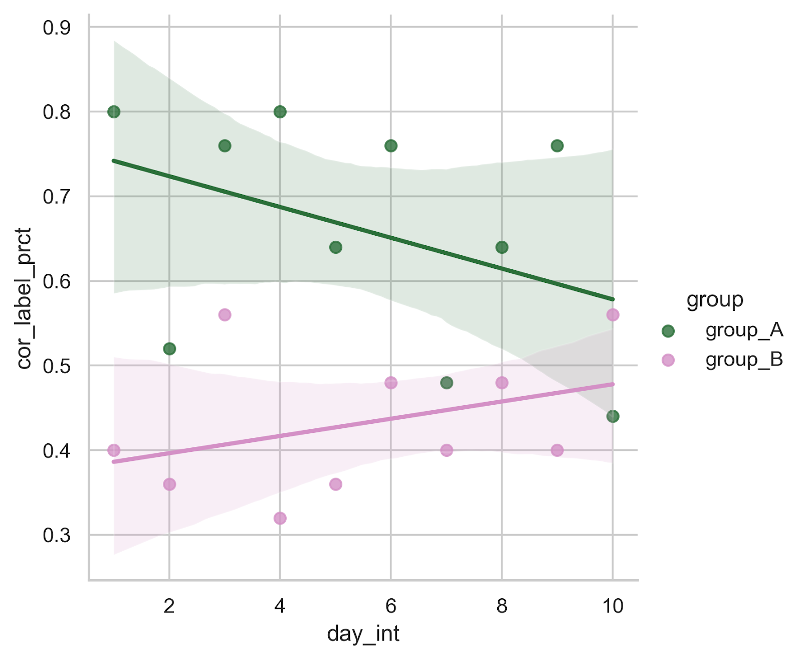}
    \caption{Trendlines of average correct label counts by time, between group A and group B participants}
    \label{fig:13}
\end{figure}

Figure \ref{fig:13} suggests that assigning more weight toward “model uncertainty” (potentially causing more human uncertainty over time) had a different influence on group A and group B participants in terms of their labeling quality. Group A participants had a significant downward trend ($JT=79.5$; $p<0.001$), whereas the trend of group B was not statistically significant. The slightly improving performances of group B over successive rounds may be the result of several factors that require further discussion (section 4.7). 

\subsection{Summary of Findings}
This case study (as well as the simulation posed in section 5.3.2) evaluated the performance of our proposed EDIG AL method versus the baseline RBM AL method, in terms of model improvements and human-model team interactions. We will discuss the results below in terms of the research questions asked earlier (whether using the EDIG strategy actually works).
\newline
\newline
\textbf{RQ-2.1}: Does the selection of those query instances (to label), that the human is more likely to be confident about (but the model is uncertain about), lead to faster learning by the model? Alternatively, does consideration of human confidence confer no benefit, and is it sufficient to simply ask the human to label instances that the model is uncertain about? 
\newline
\newline
The simulation test showed that EDIG AL outperformed RBM AL after about 30\% of the iterations were completed across different datasets (using the RF algorithm), showing the generalizability of EDIG. 

Case study III provided training with real outbound anomaly detection dataset and expert knowledge for both the EDIG and RBM approaches. AL using EDIG sampling in this case outperformed AL with RBM sampling (Figure \ref{fig:8}). 

I also examined model performance achieved after all the rounds/iterations were completed in the case study (Figures \ref{fig:11} and \ref{fig:12}). There were clear differences in the performance of the different human analysts as reflected in F1 scores.
\newline
\newline
\textbf{RQ-2.2}: Does EDIG AL models lead to better:
\newline
\newline
\textbf{a)} confidence in labelling?
\newline
\newline
The expert self-reported confidence was found to be higher with EDIG sampling than with RBM sampling (Figure \ref{fig:9}, right panel) but this difference was not statistically significant at the .05 level ($U=4293.5$; $p=0.084$). Group A had relatively lower confidence levels (Figure \ref{fig:11}), showing evidence of a Dunning-Kruger effect (where more skilled people underestimate their ability and less skilled people overestimate their ability). Group A participants also had a slight positive trend (see Figure \ref{fig:12}) in the confidence-correct label ratio ($slope=0.33$) as compared to group B who showed a negative trend ($slope=-0.60$). The difference between the two correlations, as assessed using the Fisher's z test, indicated a tendency towards better calibration of confidence in the more skilled group ($z=1.94$; $p=0.052$).
\newline
\newline
\textbf{b)} label quality (as determined by agreement with ground truth labels)?
\newline
\newline
The left panel of Figure \ref{fig:9} shows that EDIG, on average, produced more accurate labels than RBM. Aside from AL model comparison, as can be seen in Figure \ref{fig:11}, group A participants were found to be able to provide more correct labels than group B participants ($U=373.0$; $p<0.001$). However, as shown in Figure \ref{fig:13}, the ability of Group A participants to detect true anomalies (decreased over time as more difficult cases were presented to them for labeling) although their performance still remained better than the labeling performance of the Group B participants. 

\subsection{Discussion}
As discussed in section 4.2 the assumptions made in this case study were:

\begin{itemize}
    \item \textbf{Assumption 1}: Model uncertainty is correlated with human uncertainty.
    \item \textbf{Assumption 2}: Human uncertainty will show a higher correlation with label quality (or correctness) than with model uncertainty (although the correlation with model uncertainty will also exist, as noted in assumption 1).
    \item \textbf{Assumption 3}: Shorter geometric distance between two instances (in a high dimensional space) implies that the amount of confidence associated with labeling on the points should be simply to the corresponding level of confidence for the other point.
    \item \textbf{Assumption 4}: Human labelers should be reasonably well calibrated in terms of how well their ratings of confidence in their labels are related to measurable uncertainty in their labels.
\end{itemize}

The results obtained in section 4.5 appear to support these assumptions, but with some qualifications. Assumption 1 and 2 were partially verified, based on the results in sections 4.5.1 and 4.5.3. The confidence (human certainty) of both groups (A and B) kept dropping over time (Figure 5-8), potentially due to the weight of the model uncertainty increasing over time. 
However, as seen in Figure \ref{fig:13}, while group A participants had better correct label counts, a slight upward trend was observed for group B participants (although it was not statistically significant). There could be several explanations for this phenomenon, for instance:

\begin{enumerate}
    \item Figure \ref{fig:13} shows that group B participants, in most rounds, provided less than 50\% correct labels. This indicates that group B participants might be assigning random labels for instances that were difficult for them. The slight upward trend in their labeling accuracy was not statistically significant, and their roughly 50\% accuracy in the final round was at a similar level to what would be expected with random guessing.
    \item Figure \ref{fig:14} below shows models trained by group B participants were found to be querying instances that had higher weighted model uncertainty values. The gap increased over time with an average difference being significant ($U=79775.5$; $n1=540$; $n2=360$; $p<0.001$). This may indicate that group B participants might be working with relatively more difficult queries (and getting more difficult over time). However, as suggested in above 1) group B participants might be assigning random (or close to random) labels. Thus they might be less impacted by the increased difficulties leading to the flat trend over time/iterations observed in their label accuracy (Figure \ref{fig:13}). 
\end{enumerate}

\begin{figure}[h!]
    \centering
    \includegraphics[width=0.5\linewidth]{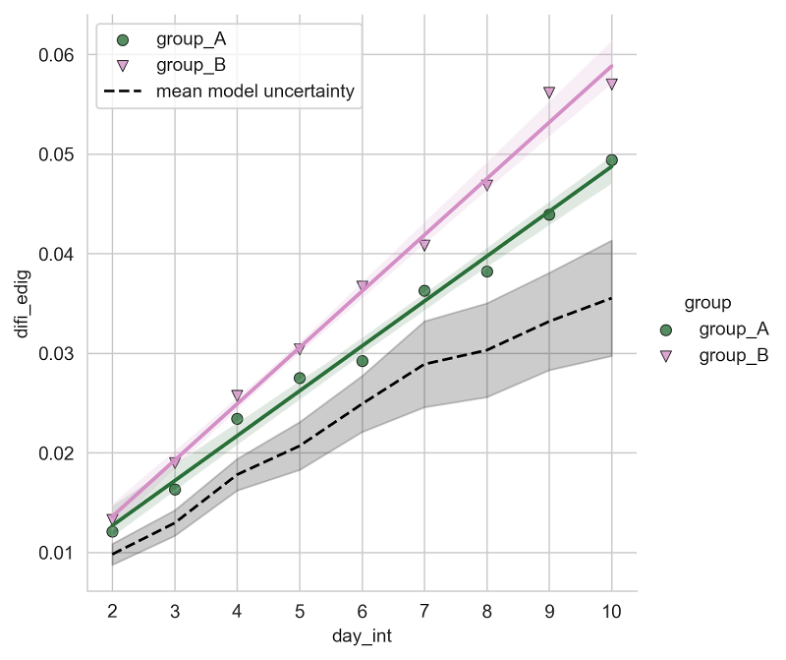}
    \caption{Average weighted model uncertainty of the queried instances in each iteration, for each group, including the mean of all remaining training instances’ weighted model uncertainty (black dash line)}
    \label{fig:14}
\end{figure}

As expected from assumption 3, confidence values for labeling EDIG instances were greater than corresponding confidence in labeling RBM instances. As can be seen in Figure \ref{fig:10}, in the first five rounds the differences were statistically significant ($U=945.0$; $n1=n2=50$; $p=0.036$). The difference of the rest rounds (last five rounds) was however not statistically significant ($U=1185.0$; $n1=n2=50$; $p=0.657$).

Assumption 4 was also found to be partially true. As can be seen in Figure \ref{fig:12}, Group A participants had a (non-significant) positive trend in terms of the increased confidence against correct label counts; whereas group B participants had a negative trend (also non-significant). If the data was not redacted and more rounds were available to be conducted with the analysts these assumptions might have been better supported by statistically significant results. However under the current limitations they were only partially true. 

Overall, the obtained results showed better performance of EDIG compared to RBM, indicating the usefulness of utilizing confidence in expert tasks to collect more correct labels. However these results also indicated the need for a training plan, or selection of more skilled labelers, generally agreeing with the results found in case study 1.

\section{Limitations}
The present research is relatively unique in looking at Cybersecurity-related AL within a large organization and we hope that it will stimulate further “realistic” research in this important area. One downside of conducting research in an industry setting is (as was true for this case study) that it is conducted in an environment with multiple constraints/limitations. For instance, due to privacy concerns and limited security clearance, some captured variables were not accessible in the present case, making the labeling task more uncertain. The limited information available in turn limited the ability of domain experts to use their knowledge fully in providing labels. 

Another limitation alluded to earlier was the number of participating human participants and their availability for multiple rounds of label. In the present study it took several weeks to get about 10 rounds of labeling data for each case study, even when only a handful labels needed to be assigned in each round. We doubt that this kind of difficulty in collecting data in this kind of setting is atypical. However, the end results of these constraints on number of iterations (rounds), size of query set per round, and number of expert participants, makes it difficult to achieve statistically significant results, when addressing some research questions.

Each of the case studies reported in this paper collected roughly 200 labels from 10 participants. While this is a relatively large sample in the context of an industry setting, it is limited in terms of its generalizability. However, it might be argued that AL will likely work differently in different settings, and thus it is more beneficial to look at the impact of confidence ratings, and different groupings of participants with small samples, but in a range of settings, rather than trying to achieve large samples but in a small number of settings. The present results are also useful in showing how AL performs in situations where available data features are constrained (due to privacy concerns). 

Human agreement in labeling as reflected in reliability scores was relatively low. This appeared to be due to some analysts having relatively low expertise with levels close to guessing for the worst performing analysts. We have no reason to believe that the overall level of expertise for analysts was particularly low relative to similar large companies. In spite of low levels of expertise for some analysts, the EDIG approach was still successful in improving AL performance, and it could be argued that it is precisely in situations where some analysts have lower expertise and thus there should be lower certainty in their labeling judgments, where an EDIG approach will be most beneficial. However, EDIG assumes that human labeling certainty can be adequately quantified and this is called into question when unskilled raters have poorly calibrated confidence judgments. In our results we found that skilled raters had better calibrated confidence ratings. Thus, the recommended solution to threats to the validity of the EDIG approach due to poor rating expertise is to either a) (preferred) select only skilled raters/analysts) or b) to train selected raters to the point where both their judgments have a high level of accuracy, and their confidence ratings are well calibrated. 

One final concern is that the studies reported did not consider the amount of time that the analysts used in judging each case during the AL process (since this information was not available). However, the efficiency of AL should ultimately be judged in terms of the tradeoff between model improvement and analyst effort expended, and thus this factor of human effort (time taken) should also be considered in future research. 

\section{Conclusions}
The findings reported here provide novel insights into the use of AL in a realistic setting, and potentially provide initial guidance on how to implement AL for cybersecurity applications within organizations.  We expect that our study design, and the findings obtained, should provide a good foundation for future researchers in this area. 

While it might seem intuitively obvious that people should not make labeling judgments in cases where they don’t have the required knowledge, the design of better AL methods is complicated by the fact that it is hard to know when people don’t know what they need to know. Confidence judgments are a potential proxy for human ability to label particular instances. However, as we showed earlier in this paper, human judges are prone to the Dunning-Kruger effect, and unskilled raters are typically overconfident in their ability to label instances. Nevertheless, as also shown in this paper, when some form of ground truth labeling is available it is possible to 1) assess the expertise of human judges and 2) assess how well calibrated their confidence judgments are. Thus it should still be possible to design valid methods of AL that take analyst confidence into account, provided that the analysts have sufficient expertise, and that their confidence judgments are sufficiently well calibrated. 

Overall, the results obtained in Study 2 demonstrate the value of using an information gain maximizing heuristic in scheduling instances to be labeled during AL. However, one caveat is that EDIG will work best in the earlier stages of AL training of ML models and that there is likely some threshold after which information gain maximization will no longer be useful. Future research will be needed to help characterize when EDIG is useful and for how long, during the AL process. The answer of when and for how long to use EDIG is likely to depend on factors such as the expertise of analysts, the quality of their confidence judgments, and the inherent uncertainty of making judgments in the particular domain under consideration.

The problem of Active Learning (AL) addressed in this paper is an example of interactive ML (iML), and more generally, a form of human-AI interaction. In this paper we demonstrated that experts can assign confidence labels that are predictive of subsequent ML prediction accuracy. Since it appears that confidence labeling is useful in guiding AL, future research should further examine the relative efficiency of AL methods that are guided by the degree of confidence that human analysts have.


\bibliographystyle{ACM-Reference-Format}
\bibliography{ref}


\begin{thebibliography}{50}


\ifx \showCODEN    \undefined \def \showCODEN     #1{\unskip}     \fi
\ifx \showDOI      \undefined \def \showDOI       #1{#1}\fi
\ifx \showISBNx    \undefined \def \showISBNx     #1{\unskip}     \fi
\ifx \showISBNxiii \undefined \def \showISBNxiii  #1{\unskip}     \fi
\ifx \showISSN     \undefined \def \showISSN      #1{\unskip}     \fi
\ifx \showLCCN     \undefined \def \showLCCN      #1{\unskip}     \fi
\ifx \shownote     \undefined \def \shownote      #1{#1}          \fi
\ifx \showarticletitle \undefined \def \showarticletitle #1{#1}   \fi
\ifx \showURL      \undefined \def \showURL       {\relax}        \fi
\providecommand\bibfield[2]{#2}
\providecommand\bibinfo[2]{#2}
\providecommand\natexlab[1]{#1}
\providecommand\showeprint[2][]{arXiv:#2}

\bibitem[Atlas et~al\mbox{.}(1989)]%
        {atlas1989training}
\bibfield{author}{\bibinfo{person}{Les Atlas}, \bibinfo{person}{David Cohn},
  {and} \bibinfo{person}{Richard Ladner}.} \bibinfo{year}{1989}\natexlab{}.
\newblock \showarticletitle{Training connectionist networks with queries and
  selective sampling}.
\newblock \bibinfo{journal}{\emph{Advances in neural information processing
  systems}}  \bibinfo{volume}{2} (\bibinfo{year}{1989}).
\newblock


\bibitem[Beckler et~al\mbox{.}(2018)]%
        {beckler2018reliability}
\bibfield{author}{\bibinfo{person}{Dylan~T Beckler}, \bibinfo{person}{Zachary~C
  Thumser}, \bibinfo{person}{Jonathon~S Schofield}, {and}
  \bibinfo{person}{Paul~D Marasco}.} \bibinfo{year}{2018}\natexlab{}.
\newblock \showarticletitle{Reliability in evaluator-based tests: using
  simulation-constructed models to determine contextually relevant agreement
  thresholds}.
\newblock \bibinfo{journal}{\emph{BMC medical research methodology}}
  \bibinfo{volume}{18} (\bibinfo{year}{2018}), \bibinfo{pages}{1--12}.
\newblock


\bibitem[Breiman(2001)]%
        {Breiman2001}
\bibfield{author}{\bibinfo{person}{L. Breiman}.}
  \bibinfo{year}{2001}\natexlab{}.
\newblock \showarticletitle{Random forests}.
\newblock  (\bibinfo{year}{2001}).
\newblock


\bibitem[Brinker(2003)]%
        {brinker2003incorporating}
\bibfield{author}{\bibinfo{person}{Klaus Brinker}.}
  \bibinfo{year}{2003}\natexlab{}.
\newblock \showarticletitle{Incorporating diversity in active learning with
  support vector machines}. In \bibinfo{booktitle}{\emph{Proceedings of the
  20th international conference on machine learning (ICML-03)}}.
  \bibinfo{pages}{59--66}.
\newblock


\bibitem[Carcillo et~al\mbox{.}(2018)]%
        {carcillo2018streaming}
\bibfield{author}{\bibinfo{person}{Fabrizio Carcillo},
  \bibinfo{person}{Yann-A{\"e}l Le~Borgne}, \bibinfo{person}{Olivier Caelen},
  {and} \bibinfo{person}{Gianluca Bontempi}.} \bibinfo{year}{2018}\natexlab{}.
\newblock \showarticletitle{Streaming active learning strategies for real-life
  credit card fraud detection: assessment and visualization}.
\newblock \bibinfo{journal}{\emph{International Journal of Data Science and
  Analytics}}  \bibinfo{volume}{5} (\bibinfo{year}{2018}),
  \bibinfo{pages}{285--300}.
\newblock


\bibitem[Cardoso et~al\mbox{.}(2017)]%
        {cardoso2017ranked}
\bibfield{author}{\bibinfo{person}{Thiago~NC Cardoso},
  \bibinfo{person}{Rodrigo~M Silva}, \bibinfo{person}{S{\'e}rgio Canuto},
  \bibinfo{person}{Mirella~M Moro}, {and} \bibinfo{person}{Marcos~A
  Gon{\c{c}}alves}.} \bibinfo{year}{2017}\natexlab{}.
\newblock \showarticletitle{Ranked batch-mode active learning}.
\newblock \bibinfo{journal}{\emph{Information Sciences}}  \bibinfo{volume}{379}
  (\bibinfo{year}{2017}), \bibinfo{pages}{313--337}.
\newblock


\bibitem[Chang and Lin(2011)]%
        {Chang2011}
\bibfield{author}{\bibinfo{person}{C.~C. Chang} {and} \bibinfo{person}{C.~J.
  Lin}.} \bibinfo{year}{2011}\natexlab{}.
\newblock \showarticletitle{LIBSVM: a library for support vector machines}.
\newblock  (\bibinfo{year}{2011}).
\newblock


\bibitem[Chung et~al\mbox{.}(2019)]%
        {chung2019efficient}
\bibfield{author}{\bibinfo{person}{John Joon~Young Chung},
  \bibinfo{person}{Jean~Y Song}, \bibinfo{person}{Sindhu Kutty},
  \bibinfo{person}{Sungsoo Hong}, \bibinfo{person}{Juho Kim}, {and}
  \bibinfo{person}{Walter~S Lasecki}.} \bibinfo{year}{2019}\natexlab{}.
\newblock \showarticletitle{Efficient elicitation approaches to estimate
  collective crowd answers}.
\newblock \bibinfo{journal}{\emph{Proceedings of the ACM on Human-Computer
  Interaction}} \bibinfo{volume}{3}, \bibinfo{number}{CSCW}
  (\bibinfo{year}{2019}), \bibinfo{pages}{1--25}.
\newblock


\bibitem[Chung et~al\mbox{.}(2020)]%
        {Chung2020}
\bibfield{author}{\bibinfo{person}{Mu~Huan Chung}, \bibinfo{person}{Mark
  Chignell}, \bibinfo{person}{Lu Wang}, \bibinfo{person}{Alexandra Jovicic},
  {and} \bibinfo{person}{Abhay Raman}.} \bibinfo{year}{2020}\natexlab{}.
\newblock \showarticletitle{{Interactive Machine Learning for Data Exfiltration
  Detection: Active Learning with Human Expertise}}. In
  \bibinfo{booktitle}{\emph{IEEE Transactions on Systems, Man, and Cybernetics:
  Systems}}, Vol.~\bibinfo{volume}{2020-Octob}. \bibinfo{pages}{280--287}.
\newblock
\showISBNx{9781728185262}
\showISSN{21682232}


\bibitem[Cohn et~al\mbox{.}(1994)]%
        {cohn1994improving}
\bibfield{author}{\bibinfo{person}{David Cohn}, \bibinfo{person}{Les Atlas},
  {and} \bibinfo{person}{Richard Ladner}.} \bibinfo{year}{1994}\natexlab{}.
\newblock \showarticletitle{Improving generalization with active learning}.
\newblock \bibinfo{journal}{\emph{Machine learning}}  \bibinfo{volume}{15}
  (\bibinfo{year}{1994}), \bibinfo{pages}{201--221}.
\newblock


\bibitem[Cover and Hart(1967)]%
        {Cover1967}
\bibfield{author}{\bibinfo{person}{T. Cover} {and} \bibinfo{person}{P. Hart}.}
  \bibinfo{year}{1967}\natexlab{}.
\newblock \showarticletitle{Nearest neighbor pattern classification}.
\newblock  (\bibinfo{year}{1967}).
\newblock


\bibitem[Danka and Horvath(2018)]%
        {Danka2018}
\bibfield{author}{\bibinfo{person}{T. Danka} {and} \bibinfo{person}{P.
  Horvath}.} \bibinfo{year}{2018}\natexlab{}.
\newblock  (\bibinfo{year}{2018}).
\newblock


\bibitem[Demir et~al\mbox{.}(2010)]%
        {demir2010batch}
\bibfield{author}{\bibinfo{person}{Beg{\"u}m Demir}, \bibinfo{person}{Claudio
  Persello}, {and} \bibinfo{person}{Lorenzo Bruzzone}.}
  \bibinfo{year}{2010}\natexlab{}.
\newblock \showarticletitle{Batch-mode active-learning methods for the
  interactive classification of remote sensing images}.
\newblock \bibinfo{journal}{\emph{IEEE Transactions on Geoscience and Remote
  Sensing}} \bibinfo{volume}{49}, \bibinfo{number}{3} (\bibinfo{year}{2010}),
  \bibinfo{pages}{1014--1031}.
\newblock


\bibitem[Dietterich(1998)]%
        {Dietterich1998}
\bibfield{author}{\bibinfo{person}{T.~G. Dietterich}.}
  \bibinfo{year}{1998}\natexlab{}.
\newblock \showarticletitle{Approximate statistical tests for comparing
  supervised classification learning algorithms}.
\newblock  (\bibinfo{year}{1998}).
\newblock


\bibitem[Dumitrache et~al\mbox{.}({[n.\,d.]})]%
        {dumitrache2018}
\bibfield{author}{\bibinfo{person}{A. Dumitrache}, \bibinfo{person}{L. Aroyo},
  {and} \bibinfo{person}{C.~(2018 Welty}.} \bibinfo{year}{[n.\,d.]}\natexlab{}.
\newblock \showarticletitle{June). Capturing ambiguity in crowdsourcing frame
  disambiguation}.
\newblock  (\bibinfo{year}{[n.\,d.]}).
\newblock


\bibitem[Estell{\'e}s-Arolas and Gonz{\'a}lez-Ladr{\'o}n-de Guevara(2012)]%
        {estelles2012towards}
\bibfield{author}{\bibinfo{person}{Enrique Estell{\'e}s-Arolas} {and}
  \bibinfo{person}{Fernando Gonz{\'a}lez-Ladr{\'o}n-de Guevara}.}
  \bibinfo{year}{2012}\natexlab{}.
\newblock \showarticletitle{Towards an integrated crowdsourcing definition}.
\newblock \bibinfo{journal}{\emph{Journal of Information science}}
  \bibinfo{volume}{38}, \bibinfo{number}{2} (\bibinfo{year}{2012}),
  \bibinfo{pages}{189--200}.
\newblock


\bibitem[Frank and Asuncion(1998)]%
        {Frank1998}
\bibfield{author}{\bibinfo{person}{A. Frank} {and} \bibinfo{person}{A.
  Asuncion}.} \bibinfo{year}{1998}\natexlab{}.
\newblock \showarticletitle{UCI machine learning repository}.
\newblock  (\bibinfo{year}{1998}).
\newblock


\bibitem[Guo et~al\mbox{.}(2017)]%
        {guo2017calibration}
\bibfield{author}{\bibinfo{person}{Chuan Guo}, \bibinfo{person}{Geoff Pleiss},
  \bibinfo{person}{Yu Sun}, {and} \bibinfo{person}{Kilian~Q Weinberger}.}
  \bibinfo{year}{2017}\natexlab{}.
\newblock \showarticletitle{On calibration of modern neural networks}. In
  \bibinfo{booktitle}{\emph{International conference on machine learning}}.
  PMLR, \bibinfo{pages}{1321--1330}.
\newblock


\bibitem[Guo and Schuurmans(2007)]%
        {guo2007discriminative}
\bibfield{author}{\bibinfo{person}{Yuhong Guo} {and} \bibinfo{person}{Dale
  Schuurmans}.} \bibinfo{year}{2007}\natexlab{}.
\newblock \showarticletitle{Discriminative batch mode active learning}.
\newblock \bibinfo{journal}{\emph{Advances in neural information processing
  systems}}  \bibinfo{volume}{20} (\bibinfo{year}{2007}).
\newblock


\bibitem[Gwizdka and Chignell(2007)]%
        {gwizdka200712}
\bibfield{author}{\bibinfo{person}{Jacek Gwizdka} {and} \bibinfo{person}{Mark
  Chignell}.} \bibinfo{year}{2007}\natexlab{}.
\newblock \showarticletitle{12. Individual Differences}.
\newblock \bibinfo{journal}{\emph{Personal information management}}
  (\bibinfo{year}{2007}), \bibinfo{pages}{206}.
\newblock


\bibitem[He et~al\mbox{.}(2014)]%
        {he2014active}
\bibfield{author}{\bibinfo{person}{Tianxu He}, \bibinfo{person}{Shukui Zhang},
  \bibinfo{person}{Jie Xin}, \bibinfo{person}{Pengpeng Zhao},
  \bibinfo{person}{Jian Wu}, \bibinfo{person}{Xuefeng Xian},
  \bibinfo{person}{Chunhua Li}, {and} \bibinfo{person}{Zhiming Cui}.}
  \bibinfo{year}{2014}\natexlab{}.
\newblock \showarticletitle{An active learning approach with uncertainty,
  representativeness, and diversity}.
\newblock \bibinfo{journal}{\emph{The Scientific World Journal}}
  \bibinfo{volume}{2014} (\bibinfo{year}{2014}).
\newblock


\bibitem[Hoi et~al\mbox{.}(2006a)]%
        {hoi2006large}
\bibfield{author}{\bibinfo{person}{Steven~CH Hoi}, \bibinfo{person}{Rong Jin},
  {and} \bibinfo{person}{Michael~R Lyu}.} \bibinfo{year}{2006}\natexlab{a}.
\newblock \showarticletitle{Large-scale text categorization by batch mode
  active learning}. In \bibinfo{booktitle}{\emph{Proceedings of the 15th
  international conference on World Wide Web}}. \bibinfo{pages}{633--642}.
\newblock


\bibitem[Hoi et~al\mbox{.}(2006b)]%
        {hoi2006batch}
\bibfield{author}{\bibinfo{person}{Steven~CH Hoi}, \bibinfo{person}{Rong Jin},
  \bibinfo{person}{Jianke Zhu}, {and} \bibinfo{person}{Michael~R Lyu}.}
  \bibinfo{year}{2006}\natexlab{b}.
\newblock \showarticletitle{Batch mode active learning and its application to
  medical image classification}. In \bibinfo{booktitle}{\emph{Proceedings of
  the 23rd international conference on Machine learning}}.
  \bibinfo{pages}{417--424}.
\newblock


\bibitem[Holub et~al\mbox{.}(2008)]%
        {holub2008entropy}
\bibfield{author}{\bibinfo{person}{Alex Holub}, \bibinfo{person}{Pietro
  Perona}, {and} \bibinfo{person}{Michael~C Burl}.}
  \bibinfo{year}{2008}\natexlab{}.
\newblock \showarticletitle{Entropy-based active learning for object
  recognition}. In \bibinfo{booktitle}{\emph{2008 IEEE Computer Society
  Conference on Computer Vision and Pattern Recognition Workshops}}. IEEE,
  \bibinfo{pages}{1--8}.
\newblock


\bibitem[John and Langley(2013)]%
        {John2013}
\bibfield{author}{\bibinfo{person}{G.~H. John} {and} \bibinfo{person}{P.
  Langley}.} \bibinfo{year}{2013}\natexlab{}.
\newblock \showarticletitle{Estimating continuous distributions in {B}ayesian
  classifiers. arXiv}.
\newblock  (\bibinfo{year}{2013}).
\newblock
\showeprint[arxiv]{1302.4964}


\bibitem[Jonckheere(1954)]%
        {jonckheere1954}
\bibfield{author}{\bibinfo{person}{A.~R. Jonckheere}.}
  \bibinfo{year}{1954}\natexlab{}.
\newblock \showarticletitle{A distribution-free k-sample test again ordered
  alternatives}.
\newblock  (\bibinfo{year}{1954}).
\newblock


\bibitem[Joshi et~al\mbox{.}(2009)]%
        {Joshi2009}
\bibfield{author}{\bibinfo{person}{A.~J. Joshi}, \bibinfo{person}{F. Porikli},
  {and} \bibinfo{person}{N.~(2009 Papanikolopoulos}.}
  \bibinfo{year}{2009}\natexlab{}.
\newblock \showarticletitle{June)}.
\newblock  (\bibinfo{year}{2009}).
\newblock


\bibitem[Jr and P.(1995)]%
        {brooks1995}
\bibfield{author}{\bibinfo{person}{Brooks Jr} {and} \bibinfo{person}{Frederick
  P.}} \bibinfo{year}{1995}\natexlab{}.
\newblock \showarticletitle{The mythical man-month: essays on software
  engineering}.
\newblock  (\bibinfo{year}{1995}).
\newblock


\bibitem[Ke et~al\mbox{.}(2017)]%
        {Ke2017}
\bibfield{author}{\bibinfo{person}{Guolin Ke}, \bibinfo{person}{Qi Meng},
  \bibinfo{person}{Thomas Finley}, \bibinfo{person}{Taifeng Wang},
  \bibinfo{person}{Wei Chen}, \bibinfo{person}{Weidong Ma},
  \bibinfo{person}{Qiwei Ye}, {and} \bibinfo{person}{Tie~Yan Liu}.}
  \bibinfo{year}{2017}\natexlab{}.
\newblock \showarticletitle{{LightGBM: A highly efficient gradient boosting
  decision tree}}.
\newblock \bibinfo{journal}{\emph{Advances in Neural Information Processing
  Systems}}  \bibinfo{volume}{2017-Decem} (\bibinfo{year}{2017}),
  \bibinfo{pages}{3147--3155}.
\newblock
\showISSN{10495258}


\bibitem[Krippendorff(2004)]%
        {Krippendorff2004}
\bibfield{author}{\bibinfo{person}{K. Krippendorff}.}
  \bibinfo{year}{2004}\natexlab{}.
\newblock \showarticletitle{Content analysis: An introduction to its
  methodology}.
\newblock  (\bibinfo{year}{2004}).
\newblock


\bibitem[Krippendorff(2011)]%
        {Krippendorff2011}
\bibfield{author}{\bibinfo{person}{K. Krippendorff}.}
  \bibinfo{year}{2011}\natexlab{}.
\newblock  (\bibinfo{year}{2011}).
\newblock


\bibitem[Lee et~al\mbox{.}(2020)]%
        {lee2020empowering}
\bibfield{author}{\bibinfo{person}{Ji-Ung Lee}, \bibinfo{person}{Christian~M
  Meyer}, {and} \bibinfo{person}{Iryna Gurevych}.}
  \bibinfo{year}{2020}\natexlab{}.
\newblock \showarticletitle{Empowering active learning to jointly optimize
  system and user demands}.
\newblock \bibinfo{journal}{\emph{arXiv preprint arXiv:2005.04470}}
  (\bibinfo{year}{2020}).
\newblock


\bibitem[Levenshtein et~al\mbox{.}(1966)]%
        {levenshtein1966binary}
\bibfield{author}{\bibinfo{person}{Vladimir~I Levenshtein} {et~al\mbox{.}}}
  \bibinfo{year}{1966}\natexlab{}.
\newblock \showarticletitle{Binary codes capable of correcting deletions,
  insertions, and reversals}. In \bibinfo{booktitle}{\emph{Soviet physics
  doklady}}, Vol.~\bibinfo{volume}{10}. Soviet Union,
  \bibinfo{pages}{707--710}.
\newblock


\bibitem[Lewis(1995)]%
        {lewis1995sequential}
\bibfield{author}{\bibinfo{person}{David~D Lewis}.}
  \bibinfo{year}{1995}\natexlab{}.
\newblock \showarticletitle{A sequential algorithm for training text
  classifiers: Corrigendum and additional data}. In
  \bibinfo{booktitle}{\emph{Acm Sigir Forum}}, Vol.~\bibinfo{volume}{29}. ACM
  New York, NY, USA, \bibinfo{pages}{13--19}.
\newblock


\bibitem[M{\'e}ndez~M{\'e}ndez et~al\mbox{.}(2022)]%
        {mendez2022eliciting}
\bibfield{author}{\bibinfo{person}{Ana~Elisa M{\'e}ndez~M{\'e}ndez},
  \bibinfo{person}{Mark Cartwright}, \bibinfo{person}{Juan~Pablo Bello}, {and}
  \bibinfo{person}{Oded Nov}.} \bibinfo{year}{2022}\natexlab{}.
\newblock \showarticletitle{Eliciting Confidence for Improving Crowdsourced
  Audio Annotations}.
\newblock \bibinfo{journal}{\emph{Proceedings of the ACM on Human-Computer
  Interaction}} \bibinfo{volume}{6}, \bibinfo{number}{CSCW1}
  (\bibinfo{year}{2022}), \bibinfo{pages}{1--25}.
\newblock


\bibitem[Murtagh and Legendre(2014)]%
        {Murtagh2014}
\bibfield{author}{\bibinfo{person}{F. Murtagh} {and} \bibinfo{person}{P.
  Legendre}.} \bibinfo{year}{2014}\natexlab{}.
\newblock \showarticletitle{Ward’s hierarchical agglomerative clustering
  method: which algorithms implement Ward’s criterion?}
\newblock  (\bibinfo{year}{2014}).
\newblock


\bibitem[Müllner(2011)]%
        {Mullner2011}
\bibfield{author}{\bibinfo{person}{D. Müllner}.}
  \bibinfo{year}{2011}\natexlab{}.
\newblock \showarticletitle{Modern hierarchical, agglomerative clustering
  algorithms. arXiv}.
\newblock  (\bibinfo{year}{2011}).
\newblock
\showeprint[arxiv]{1109.2378}


\bibitem[Niculescu-Mizil and Caruana(2005)]%
        {niculescu2005predicting}
\bibfield{author}{\bibinfo{person}{Alexandru Niculescu-Mizil} {and}
  \bibinfo{person}{Rich Caruana}.} \bibinfo{year}{2005}\natexlab{}.
\newblock \showarticletitle{Predicting good probabilities with supervised
  learning}. In \bibinfo{booktitle}{\emph{Proceedings of the 22nd international
  conference on Machine learning}}. \bibinfo{pages}{625--632}.
\newblock


\bibitem[Rosner et~al\mbox{.}(2006)]%
        {rosner2006wilcoxon}
\bibfield{author}{\bibinfo{person}{Bernard Rosner}, \bibinfo{person}{Robert~J
  Glynn}, {and} \bibinfo{person}{Mei-Ling~T Lee}.}
  \bibinfo{year}{2006}\natexlab{}.
\newblock \showarticletitle{The Wilcoxon signed rank test for paired
  comparisons of clustered data}.
\newblock \bibinfo{journal}{\emph{Biometrics}} \bibinfo{volume}{62},
  \bibinfo{number}{1} (\bibinfo{year}{2006}), \bibinfo{pages}{185--192}.
\newblock


\bibitem[Saito and Rehmsmeier(2015)]%
        {saito15}
\bibfield{author}{\bibinfo{person}{T. Saito} {and} \bibinfo{person}{M.
  Rehmsmeier}.} \bibinfo{year}{2015}\natexlab{}.
\newblock \showarticletitle{The precision-recall plot is more informative than
  the ROC plot when evaluating binary classifiers on imbalanced datasets}.
\newblock  (\bibinfo{year}{2015}).
\newblock


\bibitem[Schenk and Guittard(2009)]%
        {schenk2009crowdsourcing}
\bibfield{author}{\bibinfo{person}{Eric Schenk} {and} \bibinfo{person}{Claude
  Guittard}.} \bibinfo{year}{2009}\natexlab{}.
\newblock \showarticletitle{Crowdsourcing: What can be Outsourced to the Crowd,
  and Why?}
\newblock  (\bibinfo{year}{2009}).
\newblock


\bibitem[Settles(2009)]%
        {Settles2009}
\bibfield{author}{\bibinfo{person}{Burr Settles}.}
  \bibinfo{year}{2009}\natexlab{}.
\newblock \showarticletitle{{Active learning literature survey}}.
\newblock \bibinfo{journal}{\emph{Technical Report}} (\bibinfo{year}{2009}).
\newblock


\bibitem[Settles(2011)]%
        {Settles2011}
\bibfield{author}{\bibinfo{person}{Burr Settles}.}
  \bibinfo{year}{2011}\natexlab{}.
\newblock \showarticletitle{{From theories to queries: Active learning in
  practice}}.
\newblock \bibinfo{journal}{\emph{JMLR: Workshop and Conference Proceedings
  16}}  \bibinfo{volume}{16} (\bibinfo{year}{2011}), \bibinfo{pages}{1--18}.
\newblock


\bibitem[Settles(2012)]%
        {Settles2012}
\bibfield{author}{\bibinfo{person}{B. Settles}.}
  \bibinfo{year}{2012}\natexlab{}.
\newblock \showarticletitle{Uncertainty Sampling}.
\newblock  (\bibinfo{year}{2012}).
\newblock


\bibitem[Tamkin et~al\mbox{.}(2022)]%
        {tamkin2022active}
\bibfield{author}{\bibinfo{person}{Alex Tamkin}, \bibinfo{person}{Dat Nguyen},
  \bibinfo{person}{Salil Deshpande}, \bibinfo{person}{Jesse Mu}, {and}
  \bibinfo{person}{Noah Goodman}.} \bibinfo{year}{2022}\natexlab{}.
\newblock \showarticletitle{Active learning helps pretrained models learn the
  intended task}.
\newblock \bibinfo{journal}{\emph{arXiv preprint arXiv:2204.08491}}
  (\bibinfo{year}{2022}).
\newblock


\bibitem[Terpstra(1952)]%
        {Terpstra1952}
\bibfield{author}{\bibinfo{person}{T.~J. Terpstra}.}
  \bibinfo{year}{1952}\natexlab{}.
\newblock \showarticletitle{The asymptotic normality and consistency of
  Kendall’s test against trend, when ties are present in one ranking}.
\newblock  (\bibinfo{year}{1952}).
\newblock


\bibitem[Von~Ahn et~al\mbox{.}(2003)]%
        {von2003captcha}
\bibfield{author}{\bibinfo{person}{Luis Von~Ahn}, \bibinfo{person}{Manuel
  Blum}, \bibinfo{person}{Nicholas~J Hopper}, {and} \bibinfo{person}{John
  Langford}.} \bibinfo{year}{2003}\natexlab{}.
\newblock \showarticletitle{CAPTCHA: Using hard AI problems for security}. In
  \bibinfo{booktitle}{\emph{Eurocrypt}}, Vol.~\bibinfo{volume}{2656}. Springer,
  \bibinfo{pages}{294--311}.
\newblock


\bibitem[Von~Ahn et~al\mbox{.}(2008)]%
        {von2008recaptcha}
\bibfield{author}{\bibinfo{person}{Luis Von~Ahn}, \bibinfo{person}{Benjamin
  Maurer}, \bibinfo{person}{Colin McMillen}, \bibinfo{person}{David Abraham},
  {and} \bibinfo{person}{Manuel Blum}.} \bibinfo{year}{2008}\natexlab{}.
\newblock \showarticletitle{recaptcha: Human-based character recognition via
  web security measures}.
\newblock \bibinfo{journal}{\emph{Science}} \bibinfo{volume}{321},
  \bibinfo{number}{5895} (\bibinfo{year}{2008}), \bibinfo{pages}{1465--1468}.
\newblock


\bibitem[Wilcoxon(1992)]%
        {wilcoxon1992individual}
\bibfield{author}{\bibinfo{person}{Frank Wilcoxon}.}
  \bibinfo{year}{1992}\natexlab{}.
\newblock \showarticletitle{Individual comparisons by ranking methods}.
\newblock In \bibinfo{booktitle}{\emph{Breakthroughs in Statistics: Methodology
  and Distribution}}. \bibinfo{publisher}{Springer}, \bibinfo{pages}{196--202}.
\newblock


\bibitem[Xu et~al\mbox{.}(2003)]%
        {xu2003representative}
\bibfield{author}{\bibinfo{person}{Zhao Xu}, \bibinfo{person}{Kai Yu},
  \bibinfo{person}{Volker Tresp}, \bibinfo{person}{Xiaowei Xu}, {and}
  \bibinfo{person}{Jizhi Wang}.} \bibinfo{year}{2003}\natexlab{}.
\newblock \showarticletitle{Representative sampling for text classification
  using support vector machines}. In \bibinfo{booktitle}{\emph{Advances in
  Information Retrieval: 25th European Conference on IR Research, ECIR 2003,
  Pisa, Italy, April 14--16, 2003. Proceedings 25}}. Springer,
  \bibinfo{pages}{393--407}.
\newblock


\end{thebibliography}




\end{document}